\documentclass{emulateapj}
\usepackage{lscape}
\usepackage{url,color}
\usepackage{epstopdf}
\usepackage{epsfig}

\shorttitle{WR stars in the NIR and MIR}
\shortauthors{Faherty et al.}

\begin{document}


\title{Characterizing Wolf-Rayet Stars in the Near and Mid Infrared}


\author{Jacqueline K.\ Faherty\altaffilmark{1,2,3,6}, Michael M. Shara\altaffilmark{2,6}, David Zurek\altaffilmark{2,6}, Graham Kanarek\altaffilmark{2,5,6}, Moffat, Anthony. F. J.\altaffilmark{4}}

\altaffiltext{1}{Department of Terrestrial Magnetism, Carnegie Institution of Washington 5241 Broad Branch Road NW, Washington, DC 20015, USA jfaherty@dtm.ciw.edu}
\altaffiltext{2}{Department of Astrophysics, 
American Museum of Natural History, Central Park West at 79th Street, New York, NY 10024}
\altaffiltext{3}{Hubble Fellow}
\altaffiltext{4}{Departement de Physique, Universite de Montreal, C.P. 6128, Succursale Centre-Ville, Montreal, QC H3C 3J7, Canada}
\altaffiltext{5}{Department of Astronomy, Columbia University, Columbia Astrophysics Laboratory, 550 West 120th Street, New York, NY 10027, USA}

\altaffiltext{6}{Visiting Astronomer at the Infrared Telescope Facility, which is operated by the University of Hawaii under Cooperative Agreement no. NNX-08AE38A with the National Aeronautics and Space Administration, Science Mission Directorate, Planetary Astronomy Program}

\begin{abstract}
We present refined color-color selection criteria for identifying Wolf-Rayet (WR) stars using available mid infrared (MIR) photometry from WISE in combination with near infrared (NIR) photometry from 2MASS.  Using a sample of spectrally classified objects, we find that WR stars are well distinguished from the field stellar population in the ($W1$-$W2$) vs. ($J$-$K_{s}$) color-color diagram, and further distinguished from other emission line objects such as Planetary Nebulae, Be, and Cataclysmic variable stars using a combination of NIR and MIR color constraints.   As proof of concept we applied the color constraints to a photometric sample in the Galactic plane, located WR star candidates, and present five new spectrally confirmed and classified WC (1) and WN (4) stars.  Analysis of the 0.8-5.0 $\micron$ spectral data for a subset of known, bright WC and WN stars shows that emission lines (primarily He I) extend into the 3.0 - 5.0 $\mu$m spectral region, although their strength is greatly diminished compared to the 0.8-2.5 $\mu$m region.  The WR population stands out relative to background field stars at NIR and MIR colors due to an excess continuum contribution, likely caused by free-free scattering in dense winds.  Mean photometric properties of known WRs are presented and imply that reddened late-type WN and WC sources are easier to detect than earlier-type sources at larger Galactic radii.  WISE $W3$ and $W4$ images of 10 Wolf Rayet stars show evidence of circumstellar shells linked to mass ejections from strong stellar winds.   
\end{abstract}


\section{INTRODUCTION}
Population I, classical Wolf-Rayet (WR) stars mark the final evolutionary stages of massive stellar evolution (M $>$ 25 M$_{\sun}$ for solar metallicity; \citealt{Crowther07} and references therein). The short lives of their precursors ($<$ 10 Myr) make them excellent tracers of recent star formation. The WR phase (the final few 10$^5$ yrs of a massive star's life) is largely categorized by ferocious winds and high mass-loss rates of $\dot M$ $\sim$ 10$^{-5}$ - 10$^{-4}$ M$_\sun$ yr$^{-1}$ (see \citealt{Crowther07}).  The observational signatures of these dense, fast winds are spectra dominated by strong, velocity-broadened emission lines that may change through the short WR lifetime.   As such, they fall into specific categories based on dominant emission line elements representing distinct evolutionary phases: from nitrogen rich (subtype WN) into carbon rich (subtype WC) and finally oxygen rich (subtype WO; although this is debateably just a hotter version of the WC stage -- e.g. \citealt{Langer94}; \citealt{Norci98}).   After the WN phase, all WR stars are hydrogen deficient. Rapidly rotating WR stars are predicted to be progenitors of Type Ib and Ic core-collapse supernovae, and the sources of long-duration gamma-ray bursts.  Clearly, WR stars are critical to our understanding of star formation and stellar death processes.

While WR stars are massive, hot, and luminous, the short lives of their precursors means that they are mostly located near their nascent molecular cloud environments in the Galactic plane.  Unfortunately, this places optical surveys for Milky Way WR stars at an extreme disadvantage since severe interstellar extinction at optical wavelengths all but guarantees that sources beyond 4-5 kpc from the Sun will be essentially invisible in the optical part of the spectrum.  Yet, from the time of their discovery until the end of the 20th century, all WR stars were found through optical follow-ups. \citet{Shara99} reported that the local population was complete to $B<$ 14 mag and totaled 400 known sources in the Milky way (e.g. \citealt{van-der-Hucht01} and references there-in). However, based on an extrapolation of the local WR space density and Milky Way star and dust distribution models, one would expect roughly $\sim$6500 WR stars (\citealt{Shara99}) in total.  This would mean that the majority of WR stars lie undiscovered and optically obscured in the plane of the Galaxy. 

With the advent of infrared technology and large area surveys such as the Two Micron All Sky Survey (2MASS; \citealt{Skrutskie06}), the $Spitzer$ Galactic Legacy Infrared Mid-Plane Survey Extraordinaire data (GLIMPSE, \citealt{Benjamin03}), and the Camera Panoramique Proche InfraRouge (CPAPIR) Narrow Band Southern Galactic plane Survey (\citealt{Shara09}), a new strategy to identify WR stars has emerged that overcomes obscuration due to gas and dust.  To date more than 200 WR stars have been found using narrow and/or broad band near-infrared and mid-infrared (NIR, MIR respectively) color criteria (\citealt{Hadfield07}; \citealt{Mauerhan10,Mauerhan10a,Mauerhan10b,Mauerhan09,Mauerhan07}, \citealt{Shara09,Shara12}, Kanarek et al. in prep).  The optical classification of WR stars has been extended into the NIR (e.g. \citealt{Figer97}) and the identification of sources in well-studied clusters has provided calibrations for determining distances and extinctions to spectrally classified objects (e.g. \citealt{Crowther06}). 

With the recent data release of the Wide-Field Infrared Survey Explorer (WISE; \citealt{Wright10}), which observed the entire sky in MIR bands, we have sought to redefine the color criteria best suited for identifying WR stars.  Combining newly available WISE MIR photometry with 2MASS NIR photometry, we present color-color spaces optimized for distinguishing WR stars from the field stellar population and from other emission line objects of interest (such as planetary nebulae, cataclysmic variable, and classic Be stars).  In section 2 we review the data obtained for this analysis including the photometric sample of WR and emission line stars as well as a spectroscopic sample that we observed as part of this work.  In section 3 we define color-color spaces optimized for WR stars and in section 4 we review the results of a spectroscopic follow-up campaign to test the robustness of said constraints at identifying new Galactic sources.   In section 5, a subsample of NIR and MIR spectra for bright well studied sources are presented and used to understand the proposed photometric color spaces. In section 6 we determine the mean photometric properties of known WR stars in the MIR and NIR and propose a scheme using just ($J$-$H$) and($H$-$K_{s}$) for differentiating between WC and WN stars.  Conclusions are discussed in section 7.
	
\section{DATA\label{samples}}
Emission line stars are well-characterized by their optical and/or infrared spectra.  WR stars in particular show velocity-broadened H, He, N, C, and O lines displaying equivalent widths that can reach several hundred km s$^{-1}$.  Spectrally they are distinct; however, recent surveys following up on WR candidate stars identified using NIR to MIR photometry find that there is overlap with the broad band colors of various populations of equally interesting emission line stars (\citealt{Hadfield07}; \citealt{Mauerhan11,Mauerhan10,Mauerhan10a,Mauerhan10b,Mauerhan09,Mauerhan07}; \citealt{Shara12,Shara09}; Kanarek et al. in prep).  Among the most prominent are (1) Planetary Nebulae (PN), characterized by minimal continuum and very narrow, strong emission lines of H, He, and O, (2) Be stars, characterized by strong emission lines of H, and (3) Cataclysmic variables (CVs), characterized by broad H and He emission. In this study we focus on disentangling WR stars from these three populations using WISE-2MASS NIR and MIR color-color diagrams.  Furthermore we search for color criteria which will distinguish WR stars from background reddened stars to aid in Galactic plane searches. 

\subsection{Photometry of Emission Line Star Samples\label{emmisionline}}
 The list of known WR stars was compiled from the Galactic Census (Table 7) in \citet{Mauerhan11} in combination with the 71 new WR stars reported in \citet{Shara12}.  The list of known PN was compiled from the \citet{Kohoutek01} catalog, Be stars from the \citet{Zhang05} catalog, and CV stars from the \citet{Downes01} catalog. We cross referenced the coordinates for each PN, CV and Be star with the publicly available online catalogs of the Two micron All -Sky Survey (2MASS; \citealt{Skrutskie06}) and the Wide Field Infrared Survey (WISE; \citealt{Wright10}). We demanded a better-than 2$\arcsec$ positional match, and thereby acquired 2MASS $J,H,K_{s}$ and WISE $W1,W2,W3,W4$ photometry for each star.   We followed the online WISE All-sky data release explanatory supplement\footnote{http://wise2.ipac.caltech.edu/docs/release/allsky/expsup/index.html} which suggested that saturation became a problem at ($W1,W2,W3,W4$) $<$ (2.0, 1.5, -3.0, -4.0) respectively.  Furthermore we applied similar constraints from the 2MASS explanatory supplement that suggested saturation became a problem at ($JHK_{s}$) $<$ (5.5, 5.0, 4.5).  As source confusion and background noise in the MIR can also cause spurious detections, we also required the SNR flag for each WISE band $>$ 3 and,  following the suggestions of WISE studies such as \citet{Thompson13} and \citet{Kirkpatrick11}, we required each band to have at least seven individual images with SNR $>$ 3 in the combined Atlas image (see Table~\ref{table:addconstraints} for the full list of photometric restrictions on the sample).  In total we found 287 WR, 565 PN, 591 Be, and 137 CV stars detected in both 2MASS and WISE, with photometry that we consider to be reliable in all required bands.

In addition to comparing the properties of WR stars to emission line objects, we also compared them to Galactic plane field stars. To do so we queried the WISE catalog in a 7200$\arcsec$ square box around the arbitrary Galactic plane position 17:58:16.78 -22:52:50.9.  The WISE catalog automatically cross-identifies its stars with those in 2MASS. We restricted the sample to the same photometric criterion detailed in Table~\ref{table:addconstraints}.  Our resultant background field sample consisted of 40,891 sources matched to both 2MASS and WISE, with photometry that we consider to be reliable in all required bands.

\subsection{IRTF Spectroscopy\label{spexdata}}
Over several nights in August 2008, July 2011, and August 2012,  we used the SpeX spectrograph mounted on the 3m NASA Infrared Telescope Facility (IRTF) to obtain NIR and/or MIR spectra of (1) candidate WR stars meeting our color criteria described below, as well as (2) a sample of 10 known WR stars with a range of subtypes. The conditions of our runs were clear with average seeing (0.6 - 1.0 $\arcsec$ at $\it{J}$).    Table~\ref{SpeX} lists the details of each observation.  For the known WR sample we operated in SXD and LXD mode with the 0.8$\arcsec$ slit aligned at the parallactic angle and obtained medium resolution ($\lambda$/$\Delta$ $\lambda$ $\sim$ 1000) NIR and MIR spectral data spanning 0.8 - 5.0 $\mu$m .  For candidate WR stars we operated in prism mode with the 0.5 or 0.8$\arcsec$ slit (depending on the seeing) aligned at the parallactic angle and obtained low resolution ($\lambda$/$\Delta$ $\lambda$ $\sim$ 100) NIR spectral data spanning 0.8 - 2.5 $\mu$m. 

In SXD and Prism modes exposure times varied from 5 to 90s with 1 co-add depending on the brightness of the target.  In LXD mode exposure times were set at 0.5s due to sky brightness with 10-30 co-adds depending on the brightness of the target.  Between 2-6 images in Prism and SXD mode and 4-16 images in LXD mode were obtained for each object in an ABBA dither pattern along the slit. An A0V star was observed after each target at a similar airmass for flux calibration and telluric correction.  Internal flat-field and Ar arc lamp exposures were acquired for pixel response and wavelength calibration, respectively.  All data were reduced using SpeXtool version 3.3 (\citealt{Vacca03}, \citealt{Cushing04}) using standard settings. 

\section{COLOR-COLOR DIAGRAMS FOR EMISSION LINE OBJECTS\label{colors}}
Broad band infrared colors span areas of the spectral energy distribution that distinguish emission line objects from each other and from the field population. As such they can be used to identify photometric candidates for spectral confirmation.  For WR stars, several research teams have been using photometry of the Galactic plane to identify strong candidates for follow-up.  \citet{Hadfield07} used mid-IR $Spitzer$ GLIMPSE data, namely the $[3.6]$, $[4.5]$, $[5.8]$, and $[8.0]$ $\mu$m bands, in combination with 2MASS $JHK_{s}$ photometry, and presented the most prominent WR color-color discriminating diagrams as ($[3.6]$-$[8.0]$) vs. ($[3.6]$-$[4.5]$) and ($K_{s}$-$[8.0]$) vs.($J$-$K_{s}$).  These were subsequently used in the studies of \citet{Mauerhan09,Mauerhan10b,Mauerhan11} to identify and then confirm new WR stars.  \citet{Messineo12} used $Spitzer$ GLIMPSE data to establish a photometric classification scheme for Galactic mass-losing evolved stars (e.g. WR, Red Supergiant, and Asymptotic Giant Branch stars) with the goal of identifying strong, new, candidates from large scale photometric surveys.   Two parameters, dubbed Q1\footnote{Defined in \citet{Messineo12} as Q1 = ($J-H$) - 1.8 x ($H-K_{s}$)} and Q2\footnote{Defined in \citet{Messineo12} as Q2 = ($J-K_{s}$) - 2.69 x ($K_{s}-[8.0]$)}, were defined to distinguish various populations.  Q1 measures the deviation from the reddening vector in the ($H$-$K_{s}$) versus ($J$-$K_{s}$) plane and Q2 measures the deviation from the reddening vector in the ($J$-$K_{s}$) versus ($K_{s}$-$[8.0]$) plane (\citealt{Messineo12}, \citealt{Negueruela07}, \citealt{Comeron05}).  As the WISE bands do not completely overlap with IRAC channels used in these previous studies, a comparative as well as an expansive analysis of these works is worthwhile.  We note that the WISE ($W1, W2, W3, W4$) photometric bands are centered at (3.4, 4.6, 12.0, 22) $\micron$ with widths of (0.66,  1.04, 5.51, 4.10) respectively.  Therefore the WISE $W1$, $W2$, and $W3$ bands are similar to IRAC $[3.6]$ (3.179-3.955 $\micron$), $[4.5]$ (3.955-5.015 $\micron$), and $[8.0]$ (6.442-9.343 $\micron$) respectively.  

\begin{figure*}[!ht]
\begin{center}
\epsscale{1.0}
\plotone{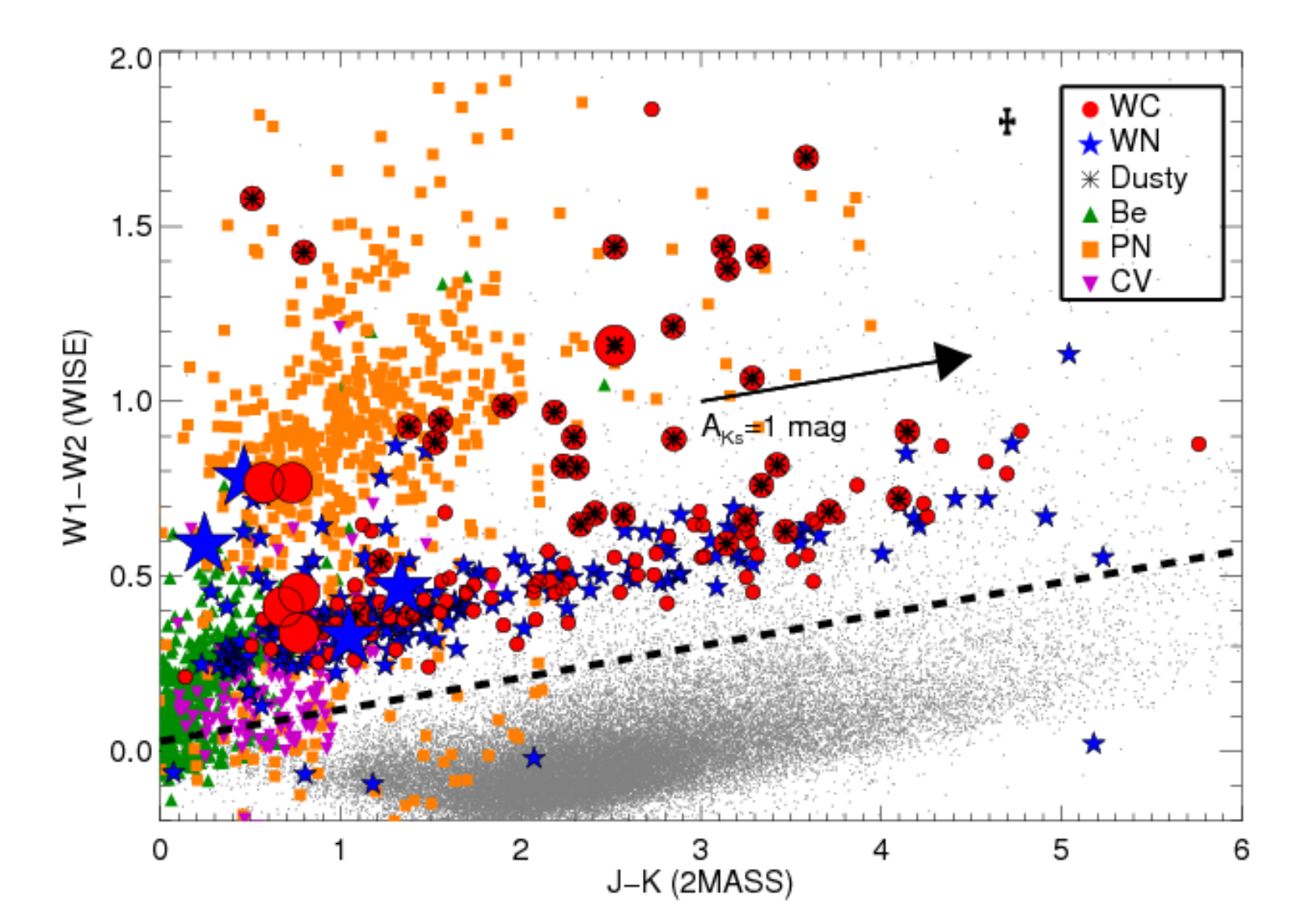}
\end{center}
\caption{The ($W1 - W2$)  vs.  ($J-K_{s}$) color-color diagram which isolates the WR stars from the vast majority of background stars (grey dots).  All sources were required to pass our photometric criterion listed in Table~\ref{table:addconstraints}.  WR stars (WC--red circles; WN--blue five-point stars) were compiled from the Galactic Census (Table 7) in \citet{Mauerhan11} in combination with the 71 new WR stars reported in \citet{Shara12}.  The list of known PN (orange squares) was compiled from the \citet{Kohoutek01} catalog, Be stars (green triangles) from the \citet{Zhang05} catalog, and CV stars (grey upside down triangle)  from the \citet{Downes01} catalog.  WR stars classified as ''Dusty" in their discovery papers are marked as asterisks. Sources presented with NIR to MIR spectra (see Table~\ref{SXD-LXD-Targets}) are emphasized as larger points.  The A$_{K_{s}}$ reddening vector is shown as well as a dashed line indicating the area optimized for WR stars. The mean photometric uncertainty for WR stars is shown to the left of the legend. } 
\label{fig:nearmidIR}
\end{figure*}

\begin{figure*}[!ht]
\begin{center}
\epsscale{1.0}
\plotone{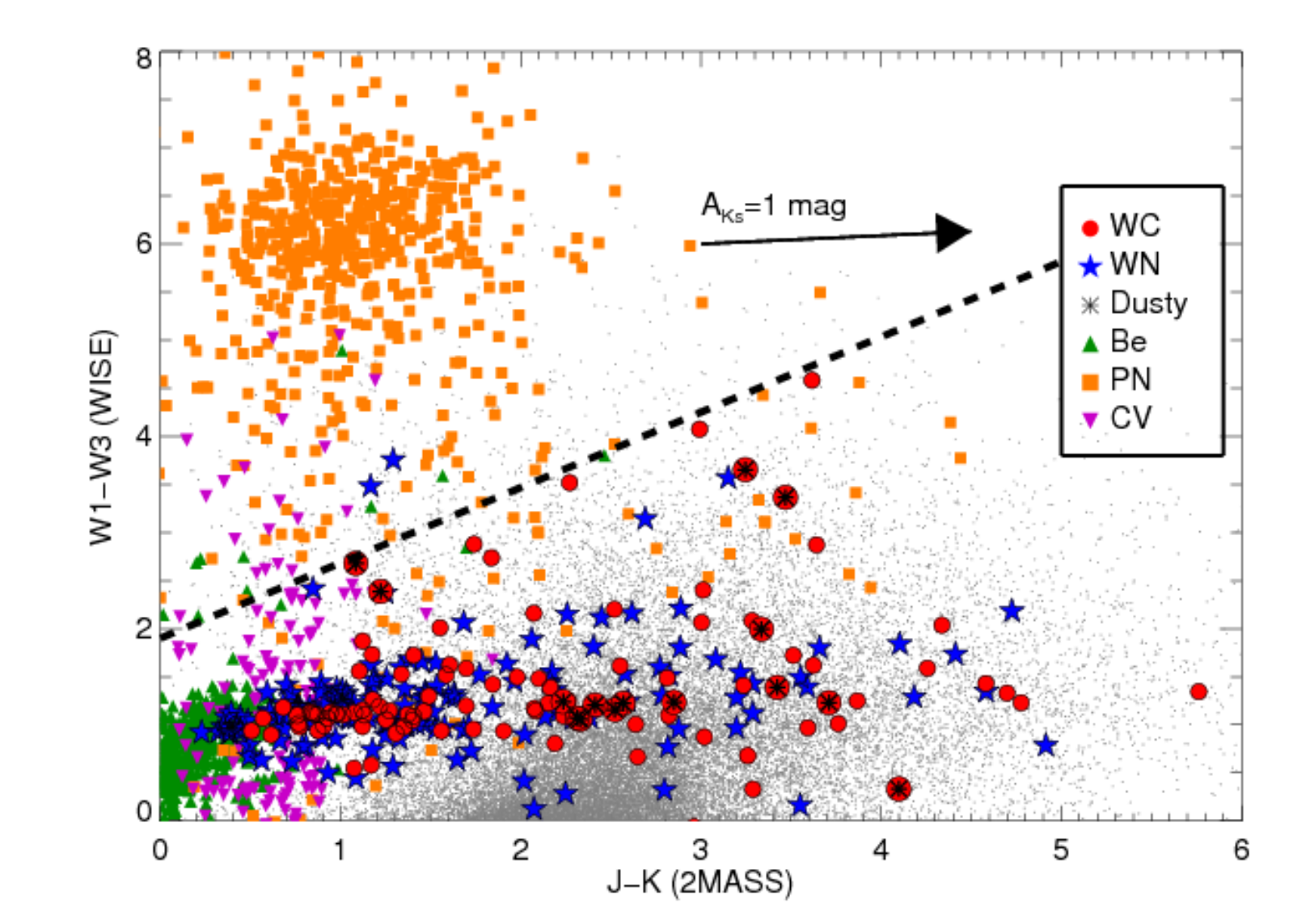}
\end{center}
\caption{The ($J-K_{s}$) vs. the ($W1$-$W3$)  color-color diagram. Symbols described in Figure~\ref{fig:nearmidIR}.}
\label{fig:W1W3}
\end{figure*}

\begin{figure*}[!ht]
\begin{center}
\epsscale{1.0}
\plotone{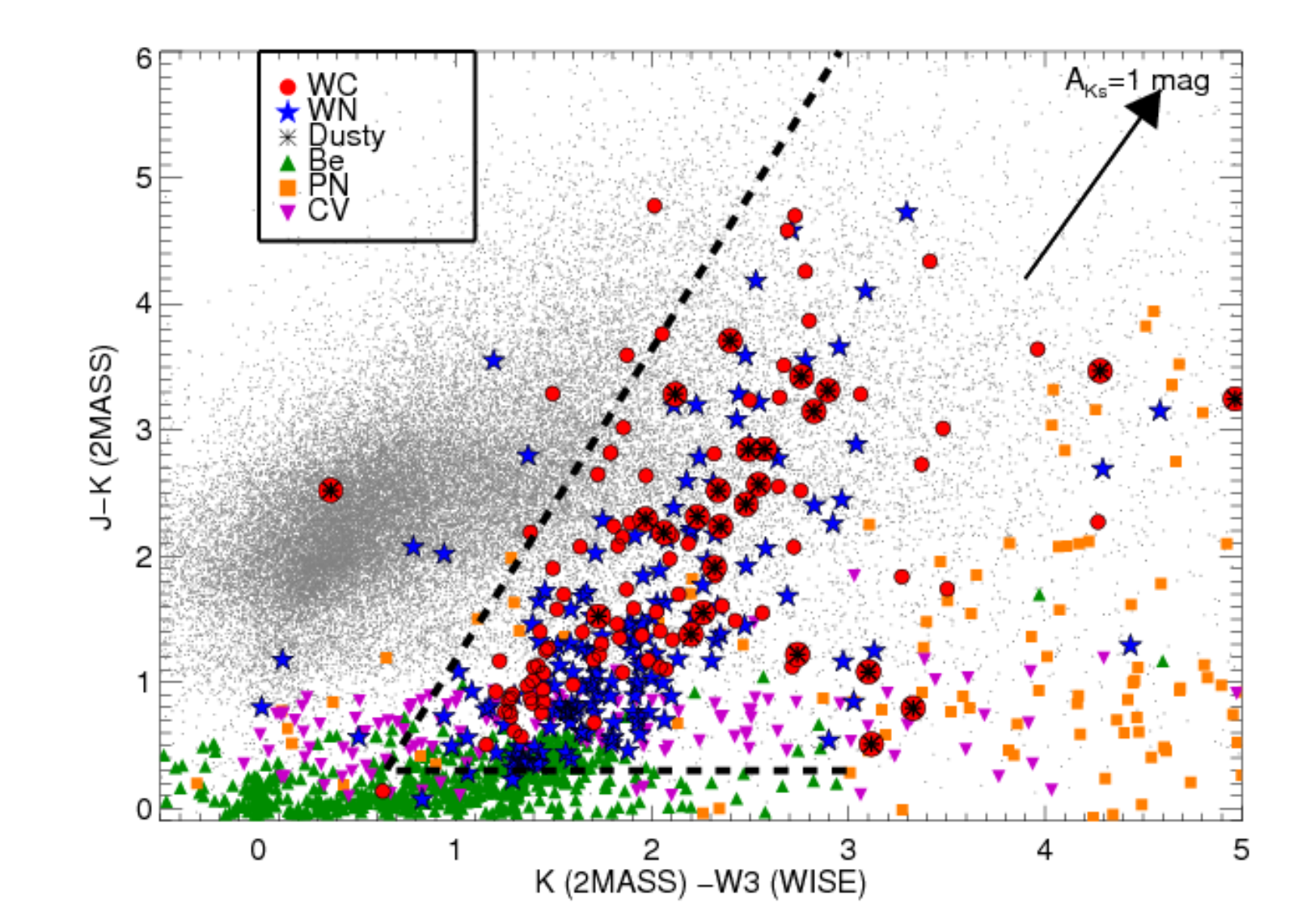}
\end{center}
\caption{The ($J-K_{s}$) vs. the ($K_{s}-W3$)  color-color diagram. Symbols described in Figure~\ref{fig:nearmidIR}.} 
\label{fig:nearmidIR2}
\end{figure*}

\begin{figure*}[!ht]
\begin{center}
\epsscale{1.0}
\plotone{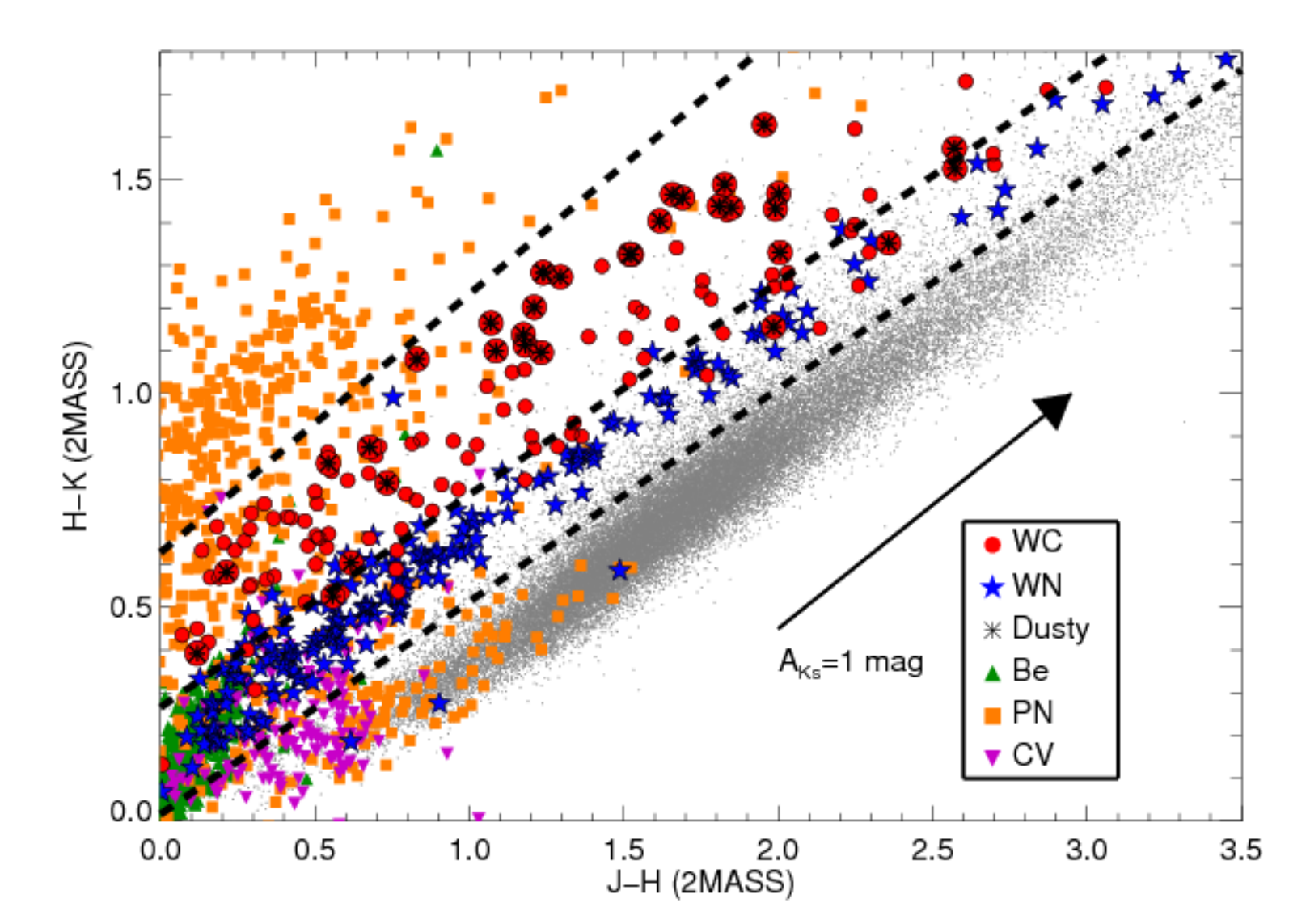}
\end{center}
\caption{The ($H-K_{s}$) vs. the ($J-H$) color-color diagram. Symbols described in Figure~\ref{fig:nearmidIR}. } 
\label{fig:nearir}
\end{figure*}

Using the full photometric suite of $J$, $H$, $K_{s}$, $W1$, $W2$, $W3$, and $W4$ bands, we found several striking color-color diagrams (some equivalent to those noted above, and some novel) that can distinguish WR stars from highly reddened or emission line objects in the Galactic plane.  Figures~\ref {fig:nearmidIR} - ~\ref{fig:nearir} show the four most informative color-color diagrams we have found.  To disentangle the effect of extinction in the scatter of each color, we overplotted  the expected 1 magnitude reddening vector in A$_{K_{s}}$ calculated from the ratios given in \citet{Indebetouw05}.  The WISE $W1$ and $W2$ bands are similar in band center and band width to the IRAC $[3.6]$ and $[4.5]$ bands,  and we therefore adopt the extinction relations for those bands (\citealt{Indebetouw05}). However the WISE $W3$ band, centered at 12.0 $\micron$, is wider than either the IRAC $[5.8]$ or $[8.0]$ bands reported.  As the IRAC extinction ratios (A$_{[4.5]}$/A$_{K}$), (A$_{[5.8]}$/A$_{K}$), and (A$_{[8.0]}$/A$_{K}$) are consistently 0.43, we make the same assumption for the WISE $W3$ band.

\begin{figure*}[!ht]
\begin{center}
\epsscale{1.2}
\includegraphics[width=.55\hsize]{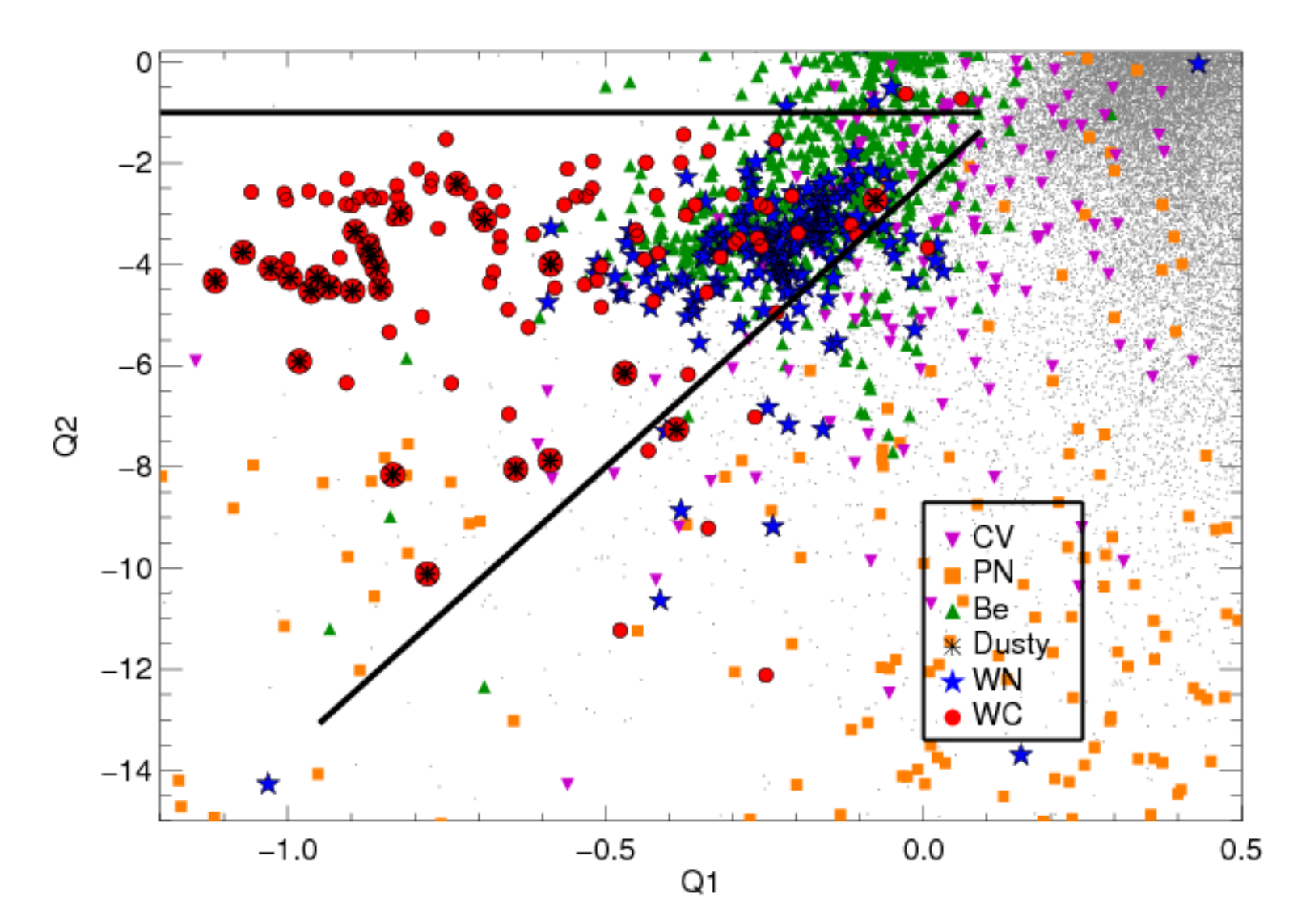}
\plottwo{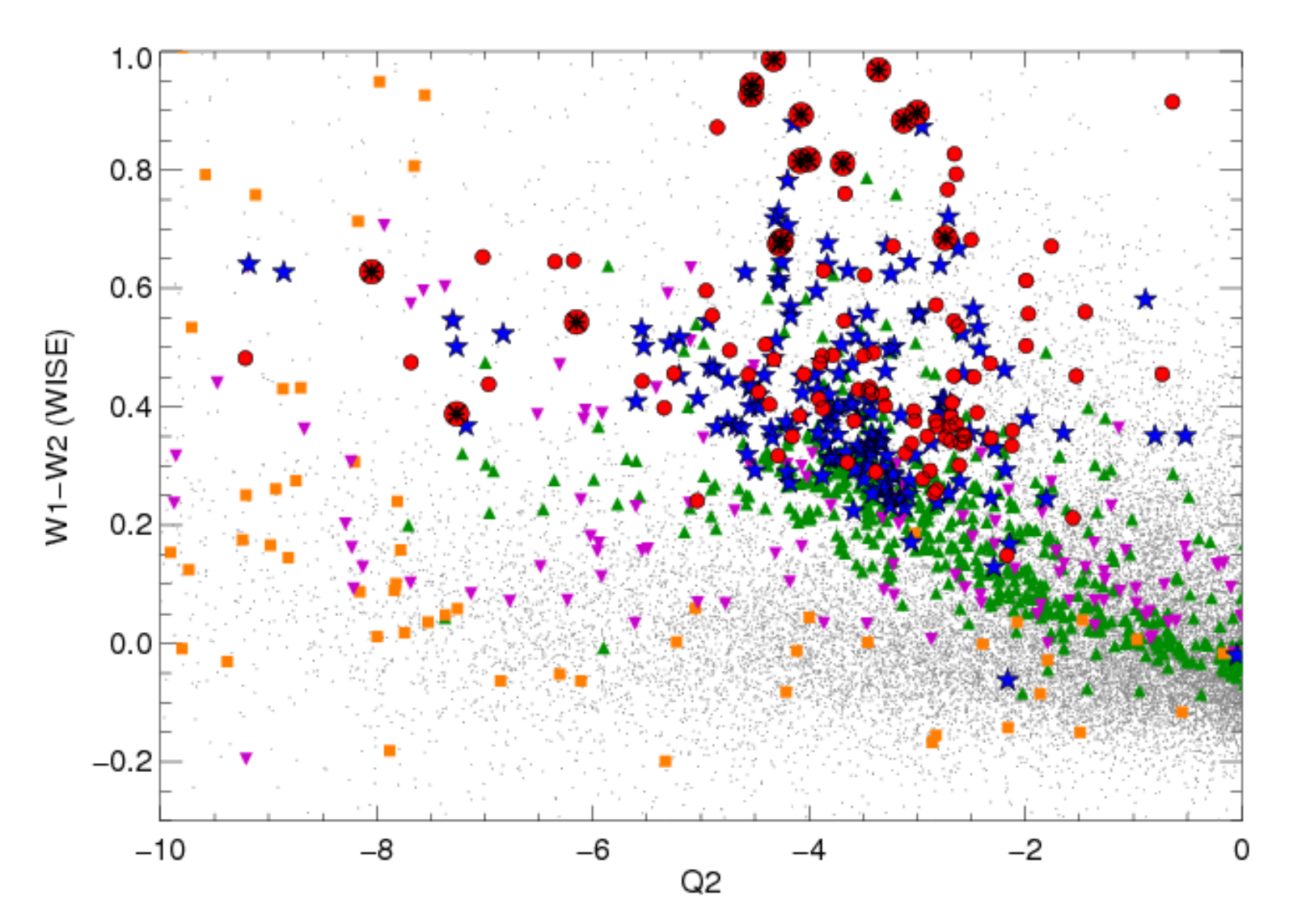}{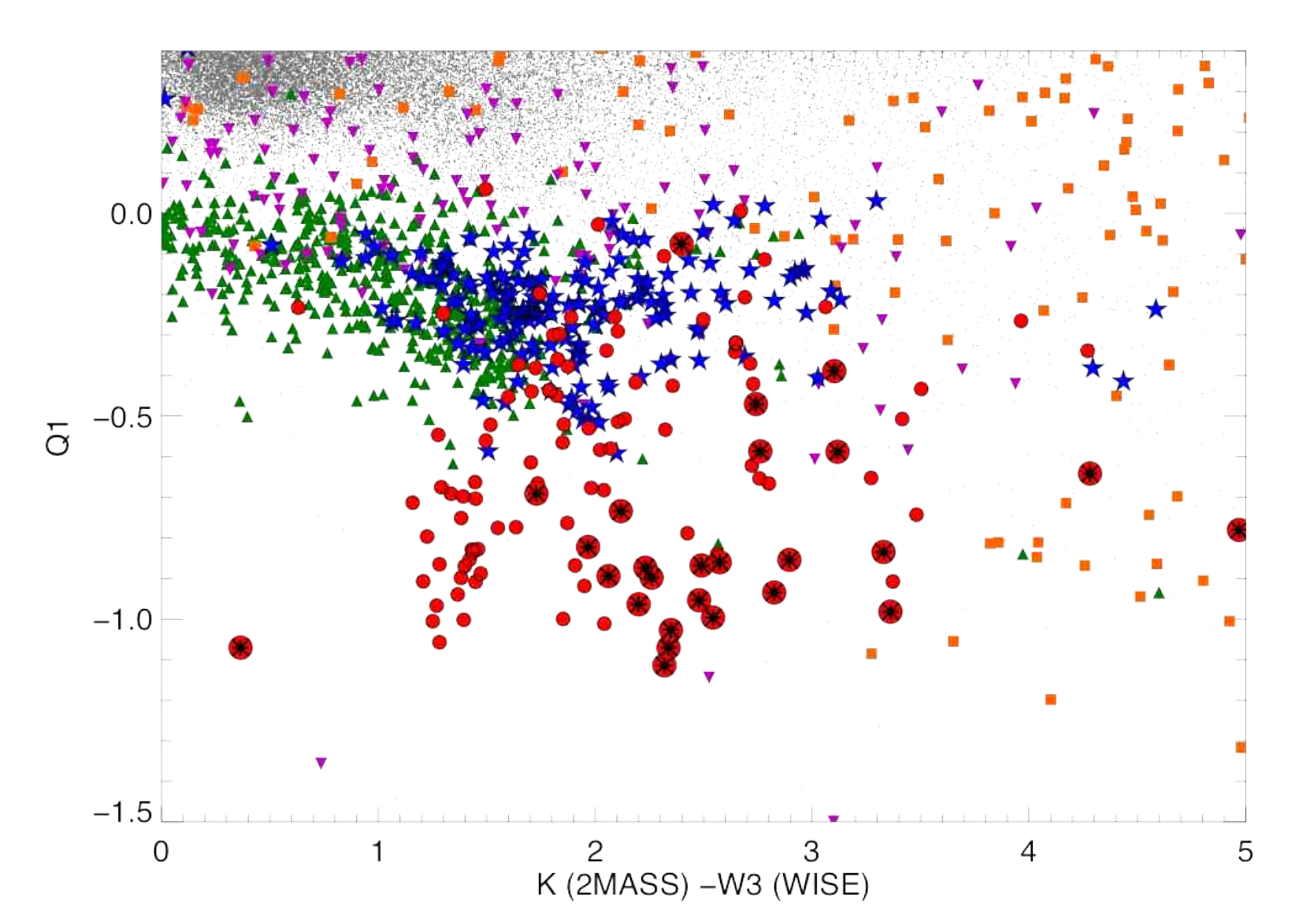}
\end{center}
\caption{ Top: The Q2 versus Q1 parameter,  Middle: The ($W1$-$W2$) vs. the Q2 parameter, Bottom:  The Q1 parameter vs. ($K_{s}$-$W3$). Q parameters presented in  \citet{Messineo12} for Spitzer channels are extrapolated for WISE photometry.  Symbols are described in Figure~\ref{fig:nearmidIR}.  We highlight the optimal WR positions described in  \citet{Messineo12} as solid lines. } 
\label{fig:q1q2}
\end{figure*}

 In Figure~\ref{fig:q1q2}, we calculate the Q1 and Q2 parameters from 2MASS and WISE data\footnote{In the case of Q2 we use the W3 band as a proxy for the IRAC $[8.0]$ channel}.  \citet{Messineo12} find that the optimal space to locate WR stars is (A) Q1 $<$ 0.1 and  (B) (11.25$\times$Q1-2.38) $<$ Q2 $<$ -1.0 (over plotted on Figure~\ref{fig:q1q2} top panel).  Primarily, this region of color-color space optimizes the location of WR stars in relation to other mass-losing evolved stars such as red super giants, and asymptotic giant branch stars. We find that  this photometric restriction (with WISE $W3$ as a proxy for IRAC [8.0] in the Q2 parameter) eliminates 98\% of the Galactic plane background sample and 72\% of the PN, however it does not distinguish the CV and Be stars (see Table~\ref{table:colorspace}). As suggested by \citet{Messineo12}, this overlap is difficult to eliminate since the photometry for each population is dominated by free-free emitters.   

\begin{figure*}[!ht]
\begin{center}
\epsscale{1.0}
\plotone{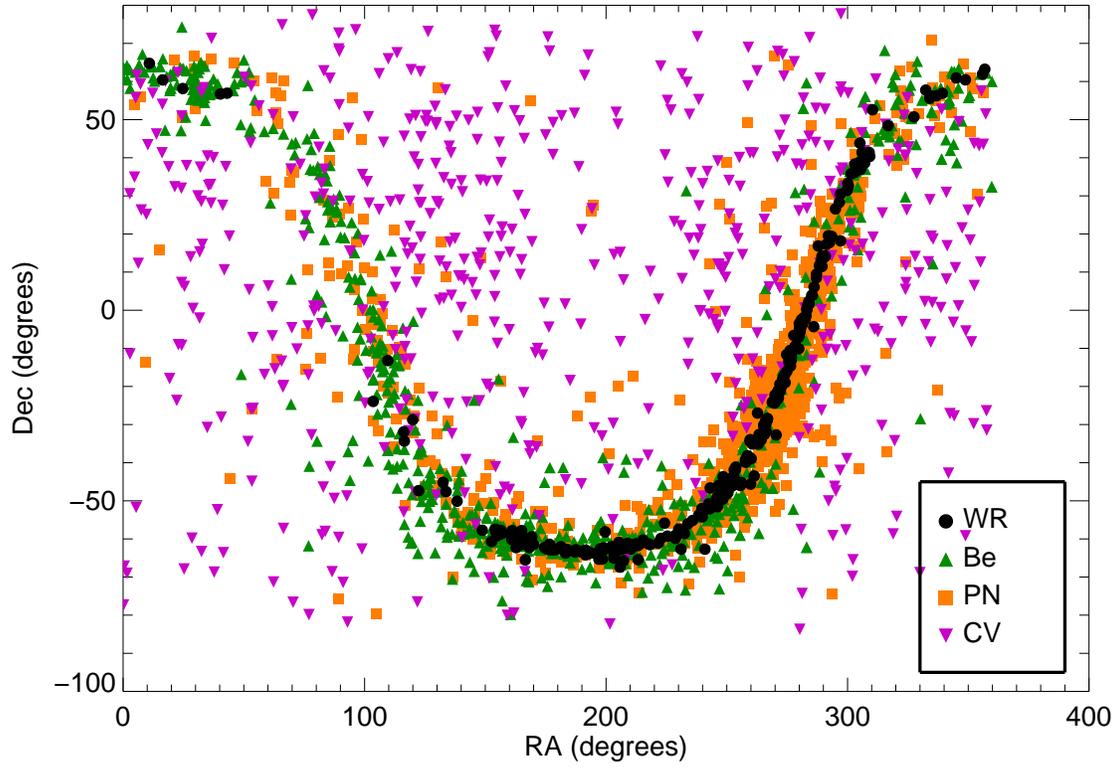}
\end{center}
\caption{The RA and DEC distribution of Galactic WR (black filled circle), Be (green triangle), CV stars (grey upside down triangle), and PN (orange squares) used in the analysis of this work.  WR stars tightly follow the Galactic center while other emission line populations spread throughout the plane.} 
\label{fig:GP}
\end{figure*}

Figure~\ref{fig:nearmidIR} demonstrates that known WRs separate from the sample of reddened field stars in the ($J$-$K_{s}$) versus ($W1$-$W2$) color-color space. We highlight the expected 1 magnitude reddening vector in A$_{K_{s}}$ and find that the contamination from PN, Be, and CV stars on this color-color diagram is concentrated at ($J$-$K_{s}$) $<$ 1.1 or an area with minimal reddening.  Be stars are intrinsically less luminous than WR stars, hence more difficult to locate at larger distances (e.g \citealt{Kozok85}).  PN  are found at comparable distances  therefore follow a similar spread in NIR color (e.g. \citealt{Stanghellini08}).  However both PN and Be stars are found scattered around the Galactic plane, not compactly concentrated (see Figure~\ref{fig:GP}).  As such, WRs spread across nearly 4 magnitudes of reddening in the NIR.

Moving to a longer wavelength regime one can readily isolate PN as a result of dust emission from their shed layers (\citealt{Anderson12}).  As shown in Figure~\ref{fig:W1W3}, PN are significantly redder in ($W1$-$W3$) than any of the other emission line stellar populations discussed here-in (see section~\ref{subsection:W4} for a discussion on the $W4$ photometry and color space).  On the basis of Figure~\ref{fig:W1W3}, we claim that searches for PN using the WISE database will be very successful if one investigates objects with ($W1$-$W3$) $>$ 4.0 mag (see also discussions in \citealt{Anderson12}, and \citealt{Parker12}).  

Shifting into the color regimes which most closely resemble those used in previous studies to identify WR candidates, we find that Figure~\ref{fig:nearmidIR2} (which resembles the ($K_{s}$-$[8.0]$) vs.($J$-$K_{s}$) diagram used in \citealt{Mauerhan11}) distinguishes the WR stars from the PN and appears to separate WRs from the Be stars.  However, IR spectroscopic follow-up of WR candidates identified with the equivalent color-color diagrams in IRAC and 2MASS bands found that Be stars remained a significant contaminant. This indicates that surveys to date have not probed deeply enough in the Galactic plane to demonstrate the range of colors that very distant, highly reddened, Be (or even CV ) stars might exhibit.  Moreover, as made clear by  Figure~\ref{fig:nearmidIR2}, WR stars have similar ($K_{s}$-$W3$) colors to background sources.  While  \citealt{Mauerhan11} discuss the ``sweet-spot" for WR discovery using the ($K_{s}$-$[8.0]$), our analysis suggests that WISE $W3$ band photometry is contaminated by diffuse background emission, causing confusion among reddened Galactic plane sources and emission line stars.  As such, we are far less confident in this color constraint as a distinguisher between sources.  We have labeled this color constraint as fifth priority (see Table~\ref{table:colorspace}) for WR searches using WISE band photometry to identify candidates.

\begin{figure*}[!ht]
\begin{center}
\epsscale{1.0}
\plotone{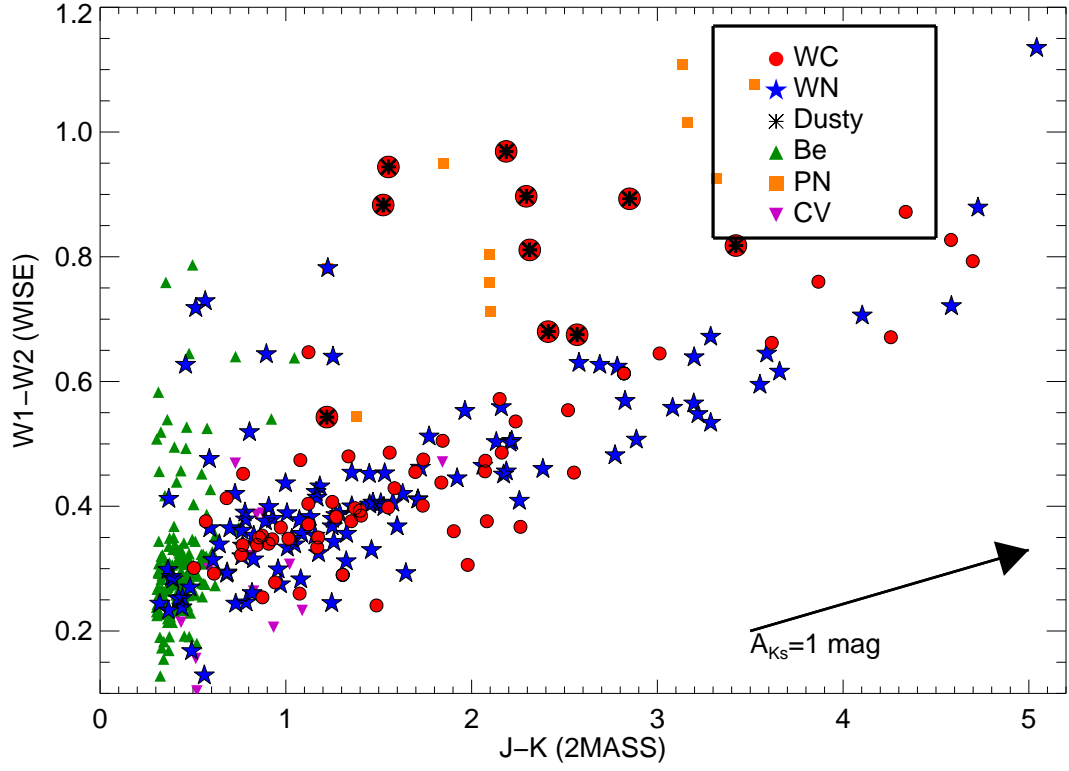}
\end{center}
\caption{The ($W1-W2$) vs. the ($J-K_{s}$) color-color diagram as in Figure~\ref{fig:nearmidIR} except with the color restrictions outlined in Table~\ref{table:colorspace} for optimizing WR star discoveries applied.} 
\label{fig:allcuts}
\end{figure*}

We suggest that a combination of constraints in both the NIR and WISE MIR will isolate WR stars from the Galactic plane background as well as distinguish them from other emission line objects. In Table~\ref{table:colorspace}, we list the WR optimized color-color spaces which are outlined by dashed lines in Figures~\ref{fig:nearmidIR} - ~\ref{fig:nearir}.  In Figure~\ref{fig:allcuts}, we apply all photometric cuts to the known populations of emission line objects discussed in this text.  After all color constraints listed in Table~\ref{table:colorspace} were applied to our photometric sample, 15 of the 565 PN remain, as do 14 of the 137 CV's, 146 of the 591 Be stars and 186 of the 287 WR stars, indicating that the constraints are effective at minimizing the contamination from emission line stars in searches for WRs.  To compare how these color-color spaces compare in their ability to isolate WR stars with that of previous Spitzer searches, we examined the 61 sources found using the ``sweet-spot" color space (in ($K_{s}$-$[8.0]$) versus ($J$-K$_{s}$)) in \citet{Mauerhan11}.  Of that sample, only 37 have 2MASS and WISE photometry suitable for this study (see the restrictions in Table~\ref{table:addconstraints}).  In a 1000 $\arcsec$ square area surrounding each of the new WR stars, we find $\sim$ 96,000 point sources in all that also have photometry suitable for analysis.  Applying all photometric cuts, we recover 70\% of the \citet{Mauerhan11} confirmed WRs and eliminate 99\% of the probable background reddened sources (leaving 550 sources in total).  

\subsection{Photometry and structure in the WISE $W4$ band\label{subsection:W4}}
\begin{figure*}[!ht]
\begin{center}
\epsscale{1.0}
\plotone{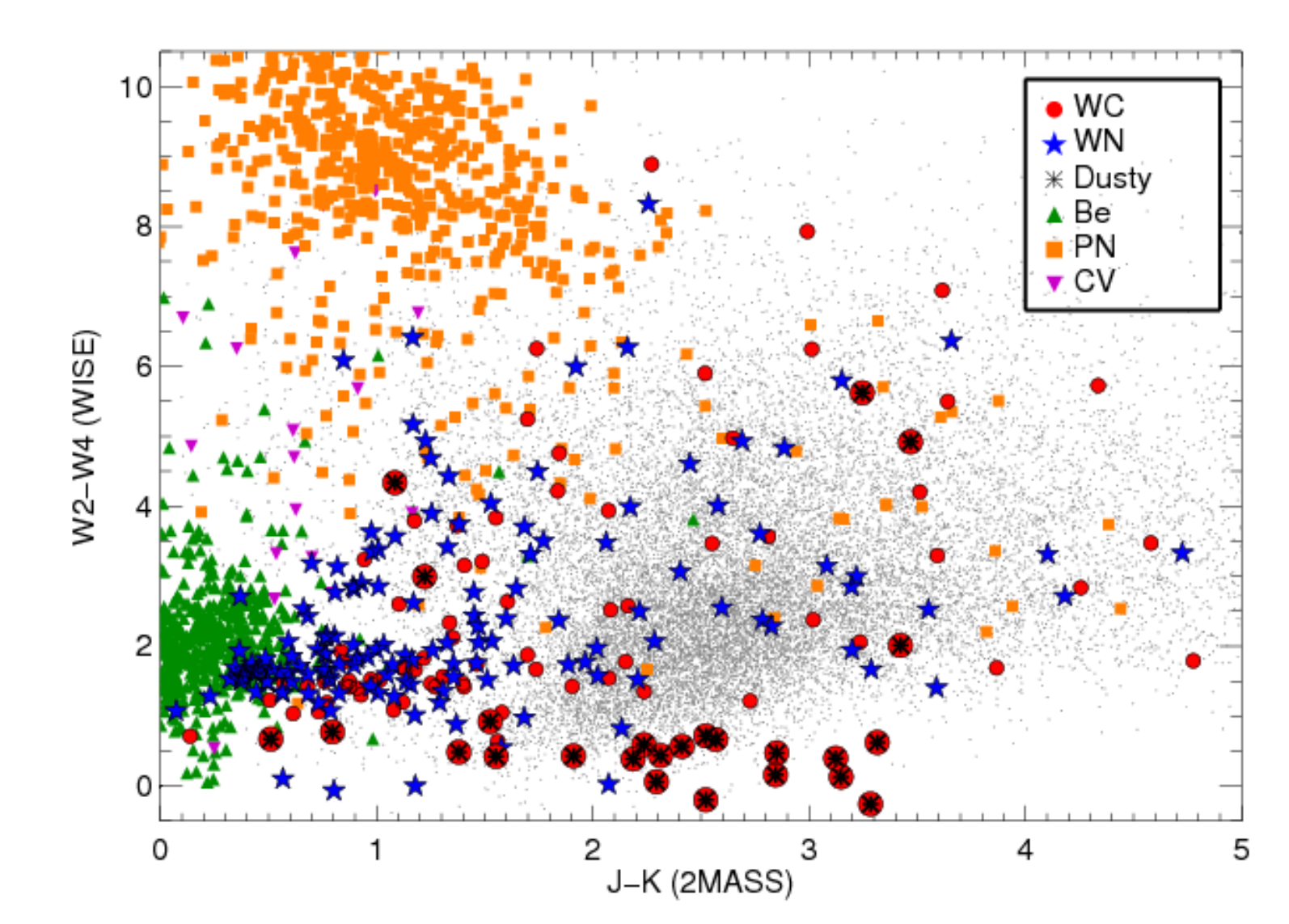}
\end{center}
\caption{The ($J-K_{s}$) vs. the ($W2$-$W4$)  color-color diagram (symbols described in Figure~\ref{fig:nearmidIR}). }
\label{fig:W4}
\end{figure*}

The 22$\micron$ (WISE $W4$) photometry for sources in crowded areas of the Galactic plane (e.g. see Figure~\ref{fig:GP} which demonstrates where the WR sources lie) are more likely to be influenced by diffuse background contamination.  Even still, we investigated whether there was an optimal color-cut that could be culled using the longest WISE wavelength band.  In Figure~\ref{fig:W4}, we show the ($J$-K$_{s}$) versus ($W2$-$W4$) color space for sources that have (1) a  non ``null" $W4$ uncertainty, (2) $W4$ mag $>$ 4, (3) $W4$ SNR $>$ 3 in at least 7 of the individual images and, (4) pass the constraints in Table~\ref{table:addconstraints}.  We find this color-color space only helps separate out the PN which were well distinguished by their red ($W1$ - $W3$) colors (see section 3 and \citealt{Anderson12}).  The Be, CV, and WR stars blend with the background field contaminants, therefore we retrieve no additional information by including $W4$ photometry in our searches for WR candidates.  

\begin{figure*}[!ht]
\begin{center}
\epsscale{.7}
\plotone{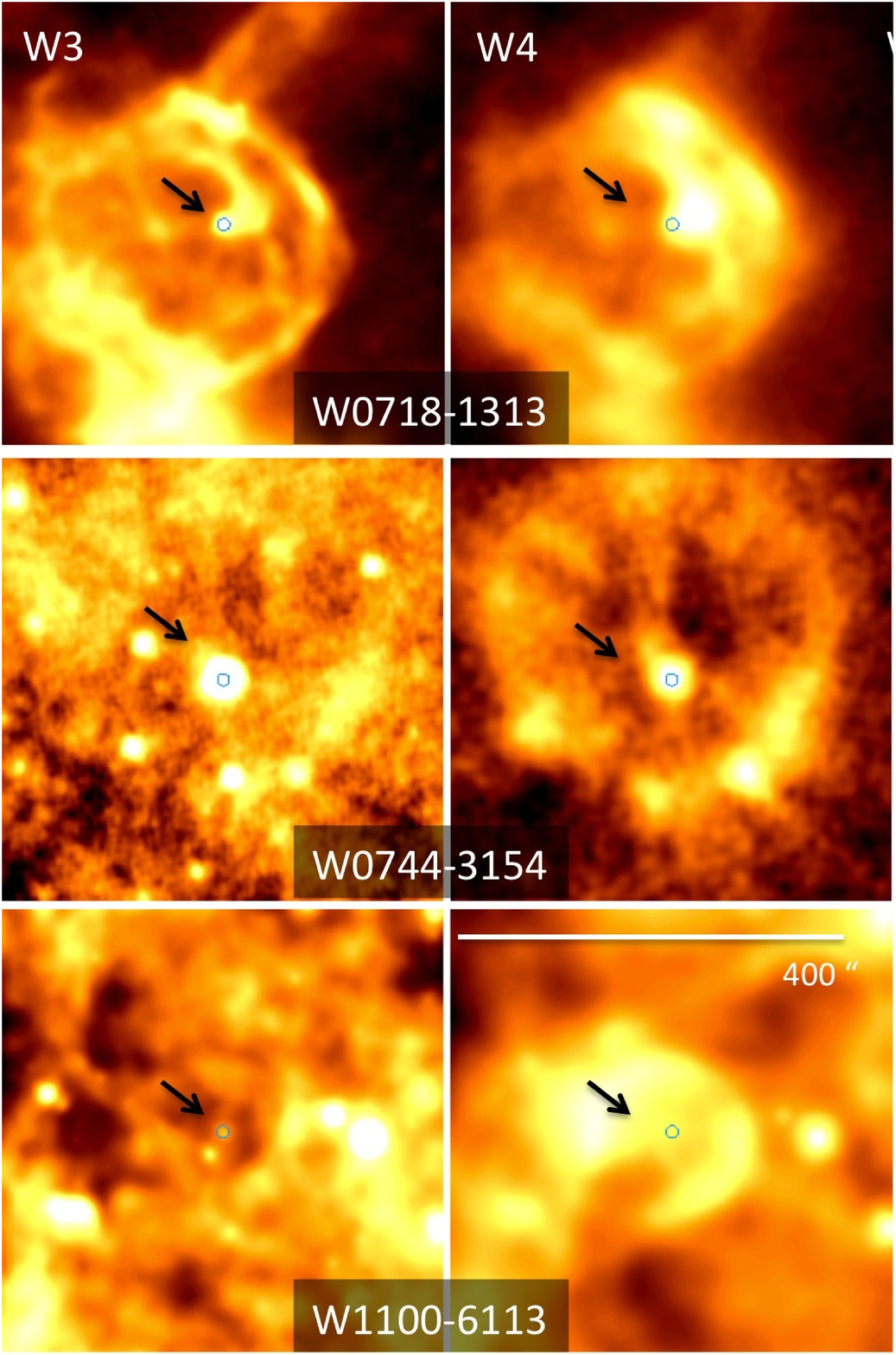}
\end{center}
\caption{The shell or bubble-like structure surrounding Wolf-Rayet stars in the WISE $W3$ and $W4$ bands.  Details on each target are listed in Table~\ref{Table:shells}.  The field of view around each is 400$\arcsec$.} 
\label{figure:W4imagesA}
\end{figure*}

\begin{figure*}[!ht]
\begin{center}
\epsscale{.7}
\plotone{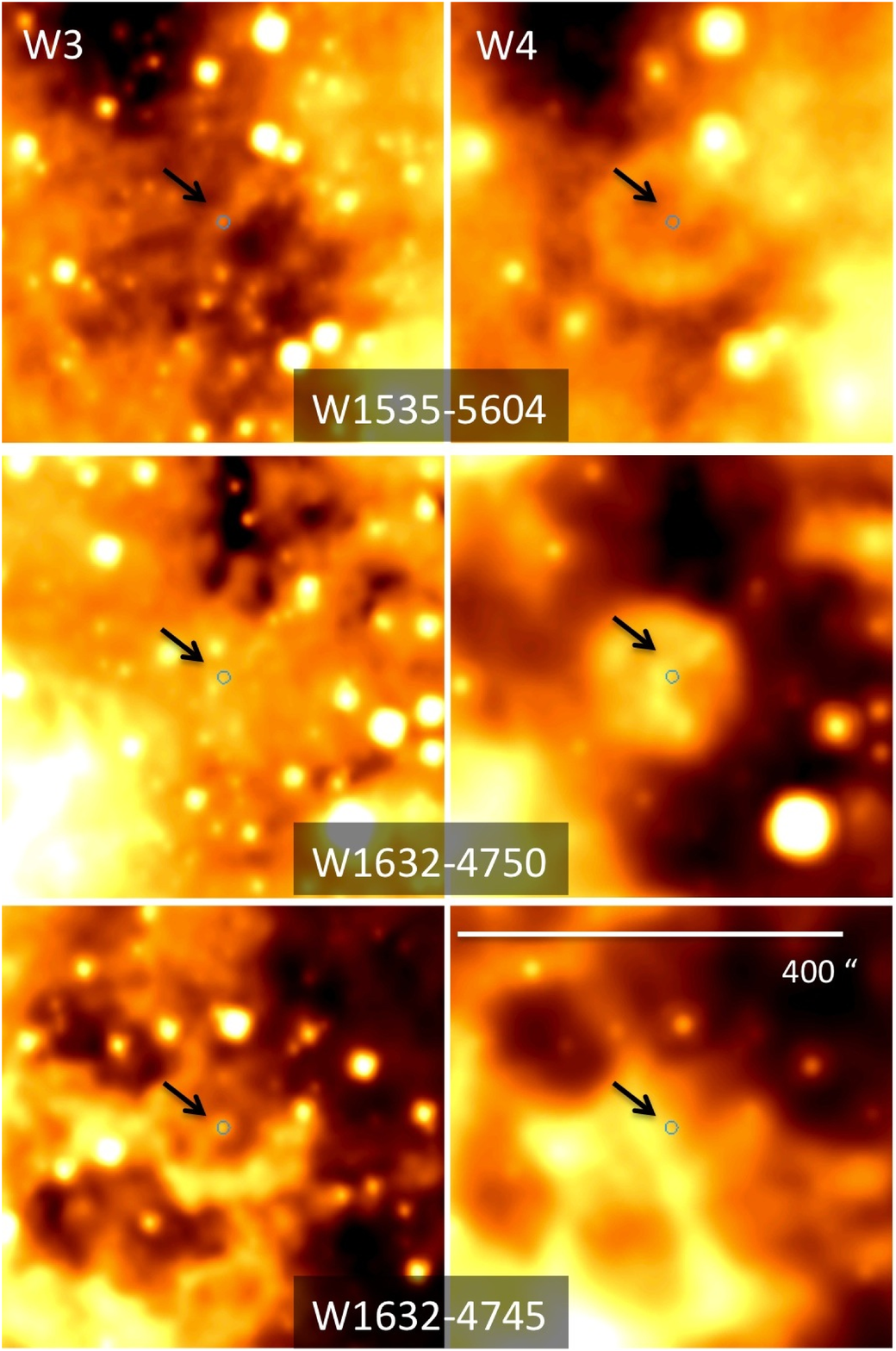}
\end{center}
\caption{The shell or bubble-like structure surrounding Wolf-Rayet stars in the WISE $W3$ and $W4$ bands.  Details on each target are listed in Table~\ref{Table:shells}.  The field of view around each is 400$\arcsec$.} 
\label{figure:W4imagesB}
\end{figure*}

\begin{figure*}[!ht]
\begin{center}
\epsscale{.7}
\plotone{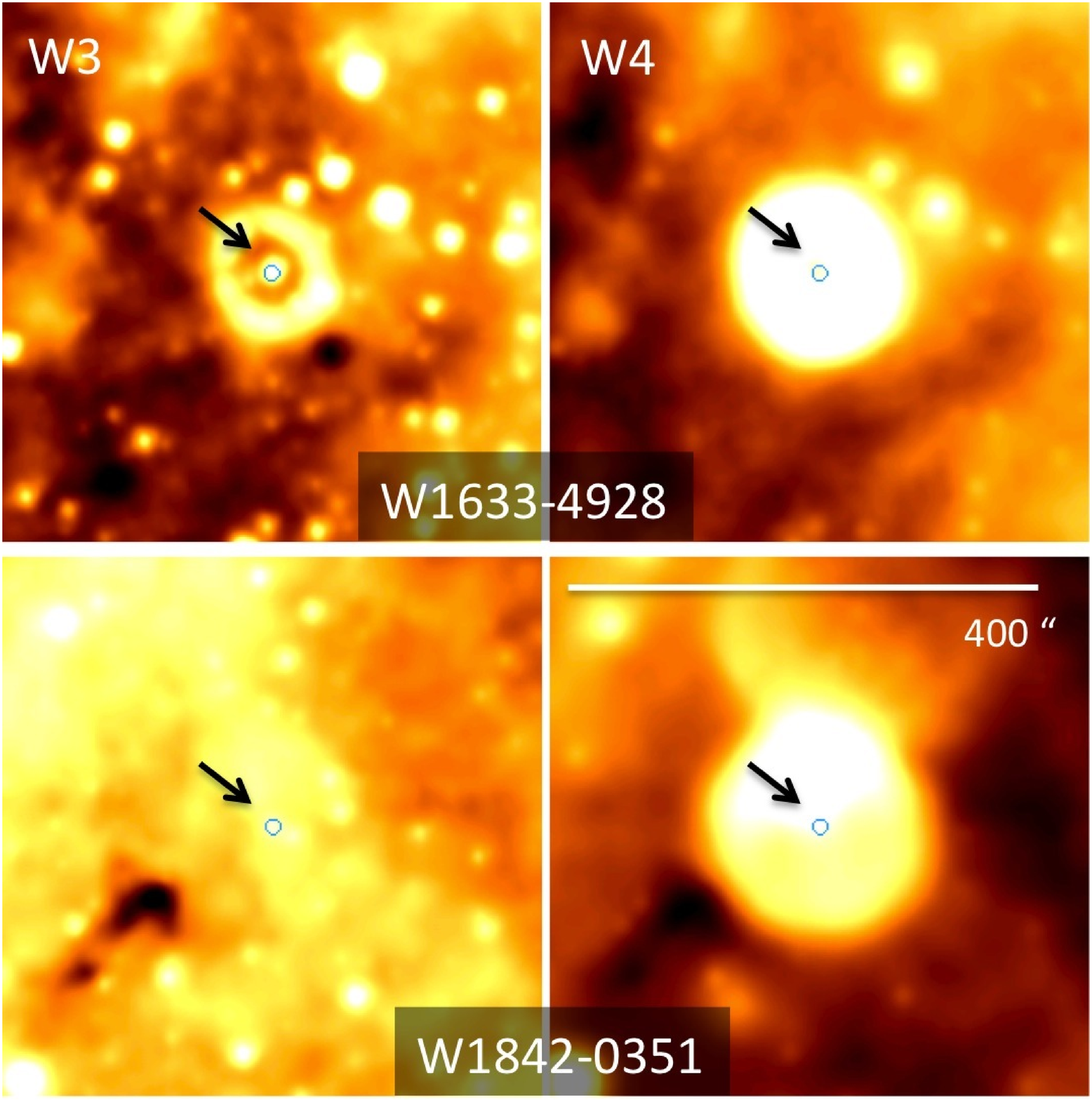}
\end{center}
\caption{The shell or bubble-like structure surrounding Wolf-Rayet stars in the WISE $W3$ and $W4$ bands.  Details on each target are listed in Table~\ref{Table:shells}.  The field of view around each is 400$\arcsec$.} 
\label{figure:W4imagesC}
\end{figure*}

\begin{figure*}[!ht]
\begin{center}
\epsscale{.7}
\plotone{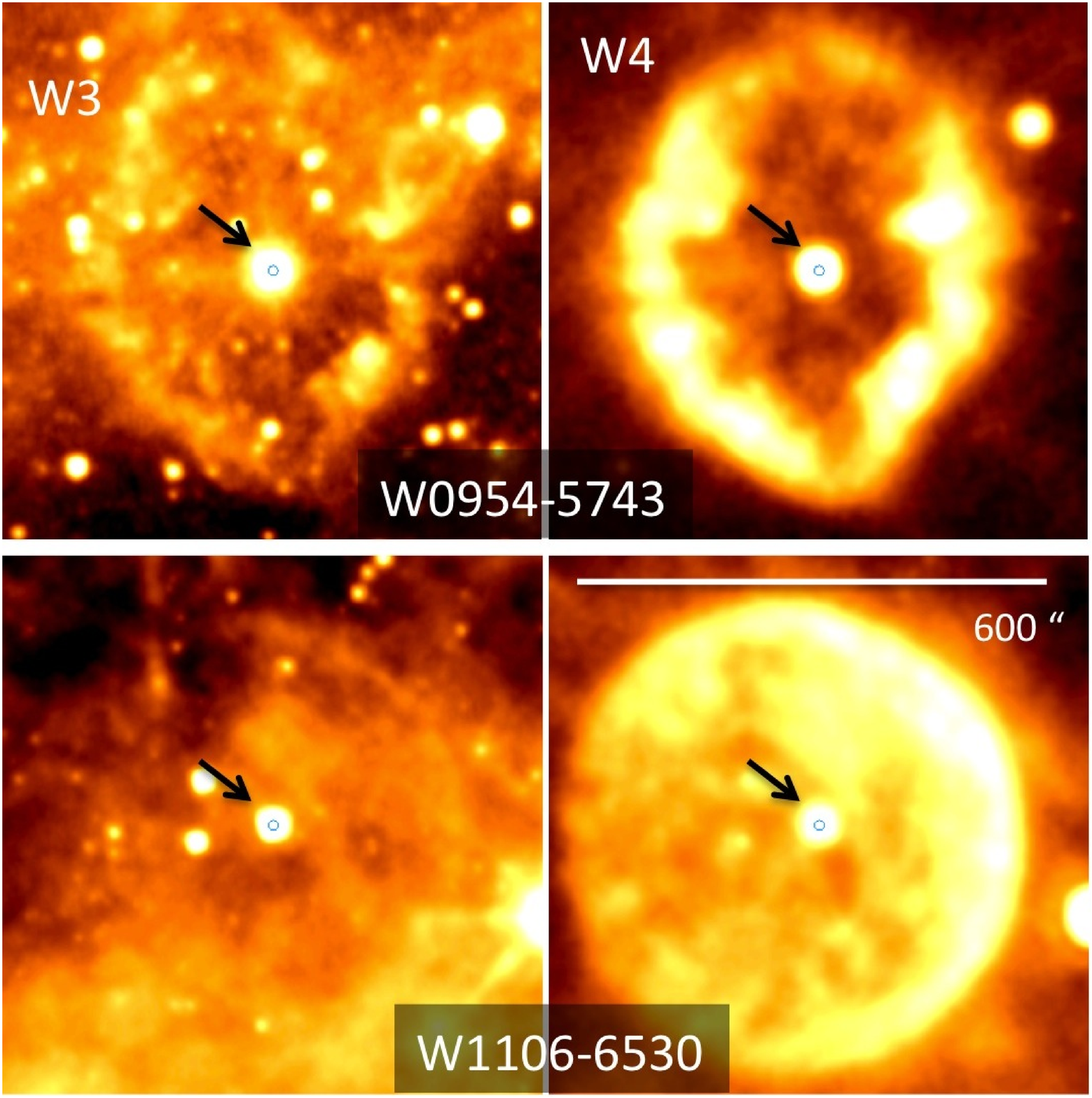}
\end{center}
\caption{The shell or bubble-like structure surrounding Wolf-Rayet stars in the WISE $W3$ and $W4$ bands.  Details on each target are listed in Table~\ref{Table:shells}.  The field of view around each is 600$\arcsec$.} 
\label{figure:W4imagesD}
\end{figure*}

Perhaps the most interesting aspect of the $W4$ photometry, is that the dusty classified WC stars have a blue ($W2$-$W4$), near zero  color compared to non-dusty sources.   We have examined the 600 $\arcsec$ area around each WR star in each WISE band image, and find no obvious  features around the dusty sources that would lead to this trend.  We did, however, uncover several interesting shell or bubble structures (see ~\citealt{Faherty13}) similar to those found using the Spitzer 24$\micron$ photometry by \citet{Wachter10}.  A number of these have been discussed in the literature, however the shell around WR1093-140LB is reported here for the first time (\citealt{Shara12}).  We list the sources that had the most striking (to our eyes) WISE $W3$ and/or $W4$ structure in Table~\ref{Table:shells} and display the gallery of shells in Figures~\ref{figure:W4imagesA} - ~\ref{figure:W4imagesD}.

\section{APPLICATION OF THE PHOTOMETRIC CONSTRAINTS\label{section:constraints}}
As a test of the ability of our color-color spaces to select WR stars, we obtained NIR spectra of a handful of strong candidates. However, as discussed in \citet{Hadfield07} and \citet{Mauerhan11} and as evidenced by the target list remaining in the vicinity of known WRs (550 from a sample of 96,000 -- see section 3 above),  supplementing broad band photometry with an additional measure of emission line activity (e.g. broad-band X-ray information as in \citealt{Mauerhan11}) greatly improves the odds of detecting a WR star.  Consequently, as an additional constraint, we selected sources in a patch of the Galactic plane that was also observed by the narrow band CPAPIR survey described in \citet{Shara09} and required that each source: (1) pass the WISE and 2MASS photometry restrictions outlined in Table~\ref{table:addconstraints}, (2) fit within the color-color spaces outlined in Table~\ref{table:colorspace},  and be (3)  $>$3$\sigma$ outliers above the continuum in one of the four narrowband CIV, HeI, HeII, and Br $\gamma$ CPAPIR filters.    As discussed above, we found that the WISE $W3$ band photometry was not as reliable as the IRAC $[8.0]$ channel, therefore we relaxed our constraint on a source having a detection in WISE $W3$.  As demonstrated by the photometry listed in Table~\ref{table:newWRs}, not all sources were detected in $W3$.   For those targets, we used only Priority 1 and 2 color constraints as listed in Table~\ref{table:colorspace}.

\begin{figure*}
\centering
\begin{tabular}{cc}
\epsfig{file=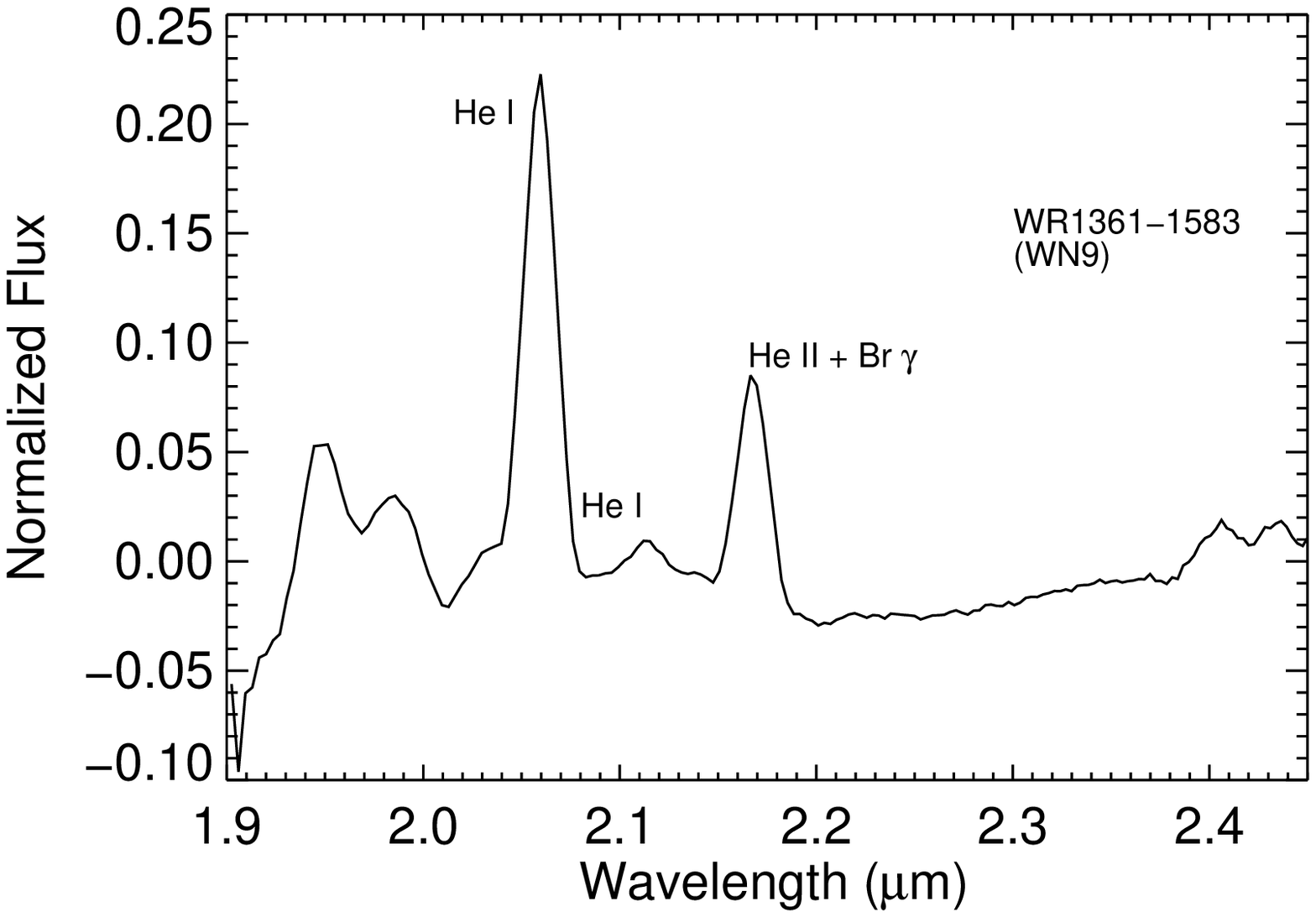,width=0.5\linewidth,clip=} &
\epsfig{file=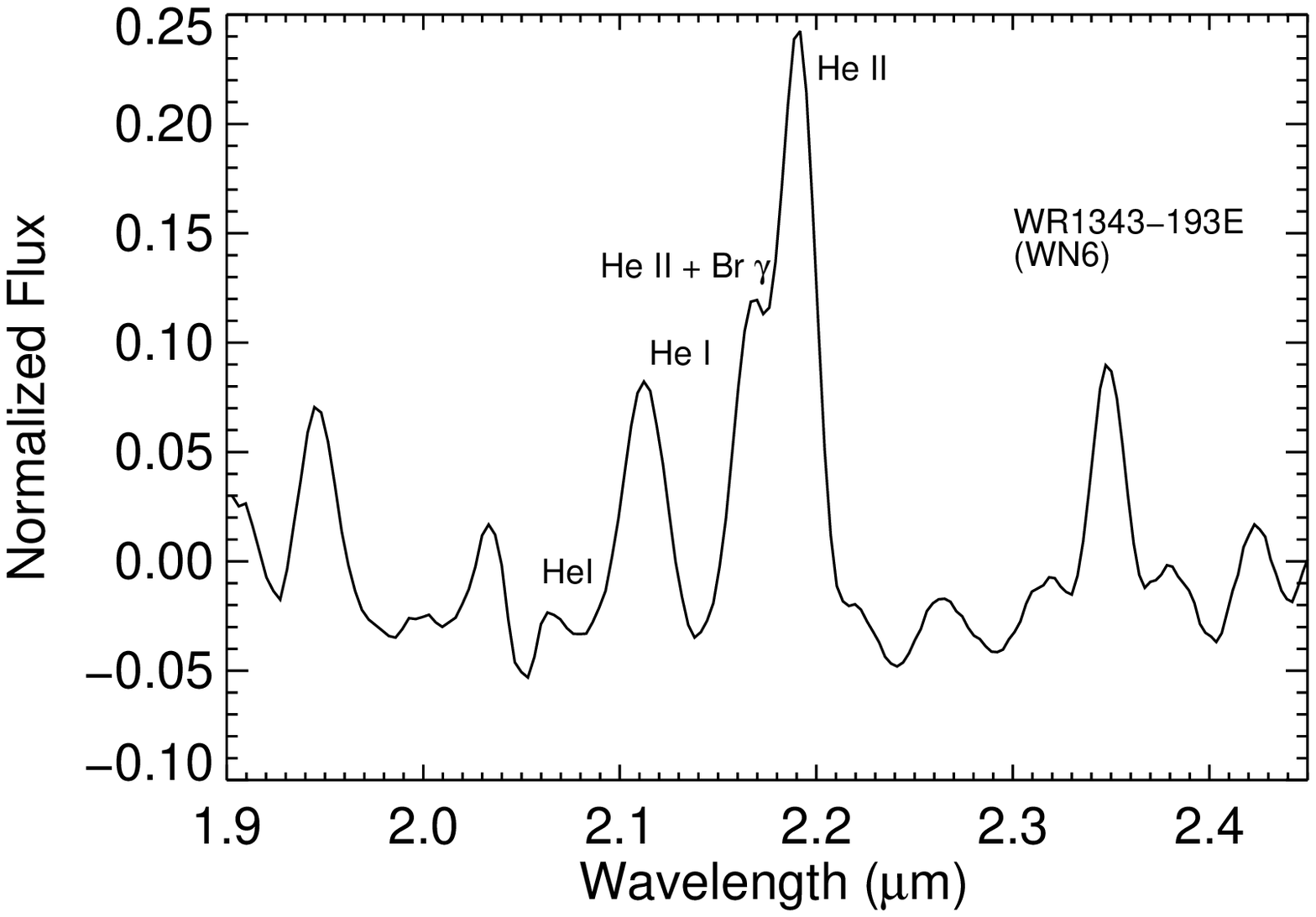,width=0.5\linewidth,clip=} \\
\epsfig{file=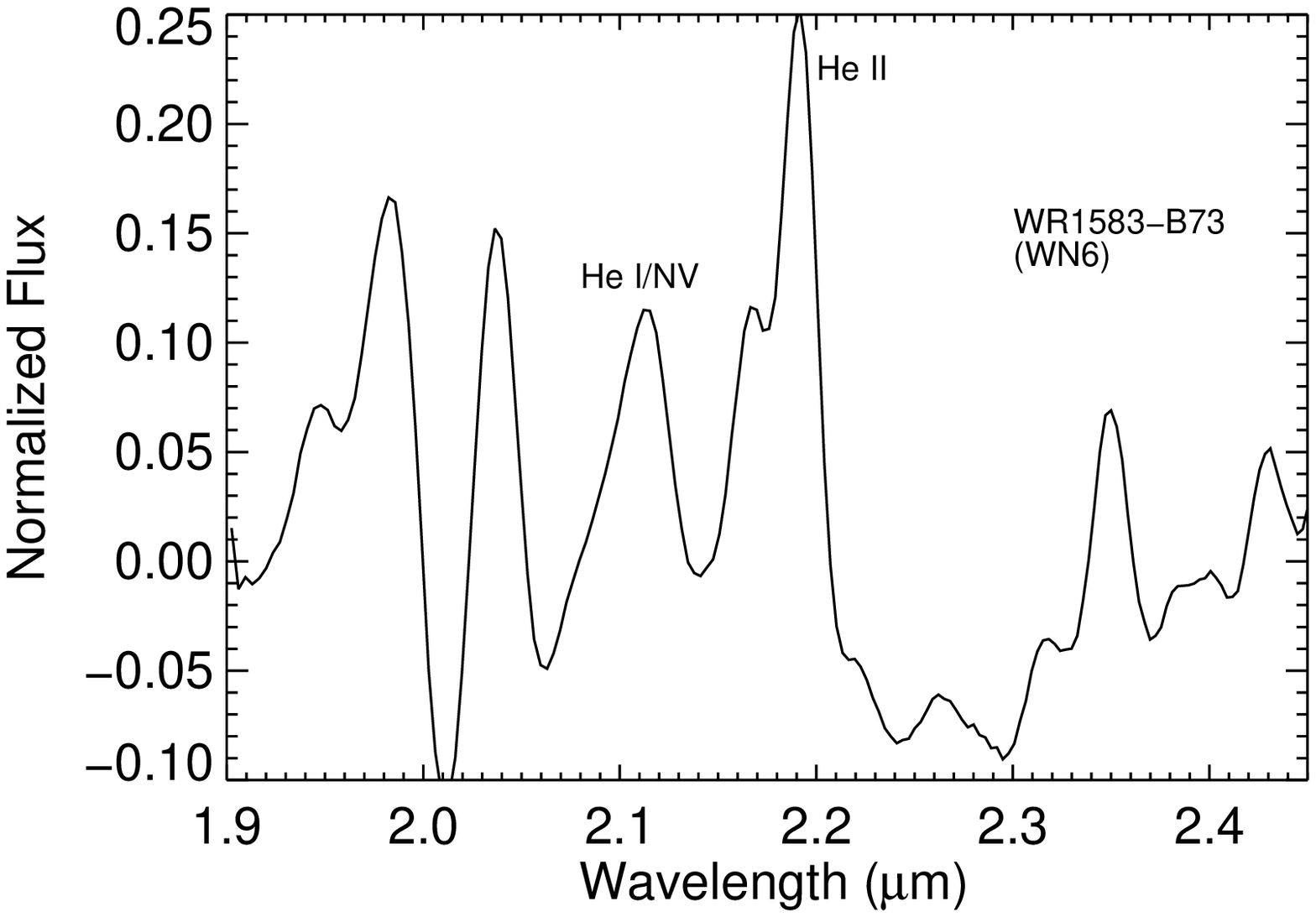,width=0.5\linewidth,clip=} &
\epsfig{file=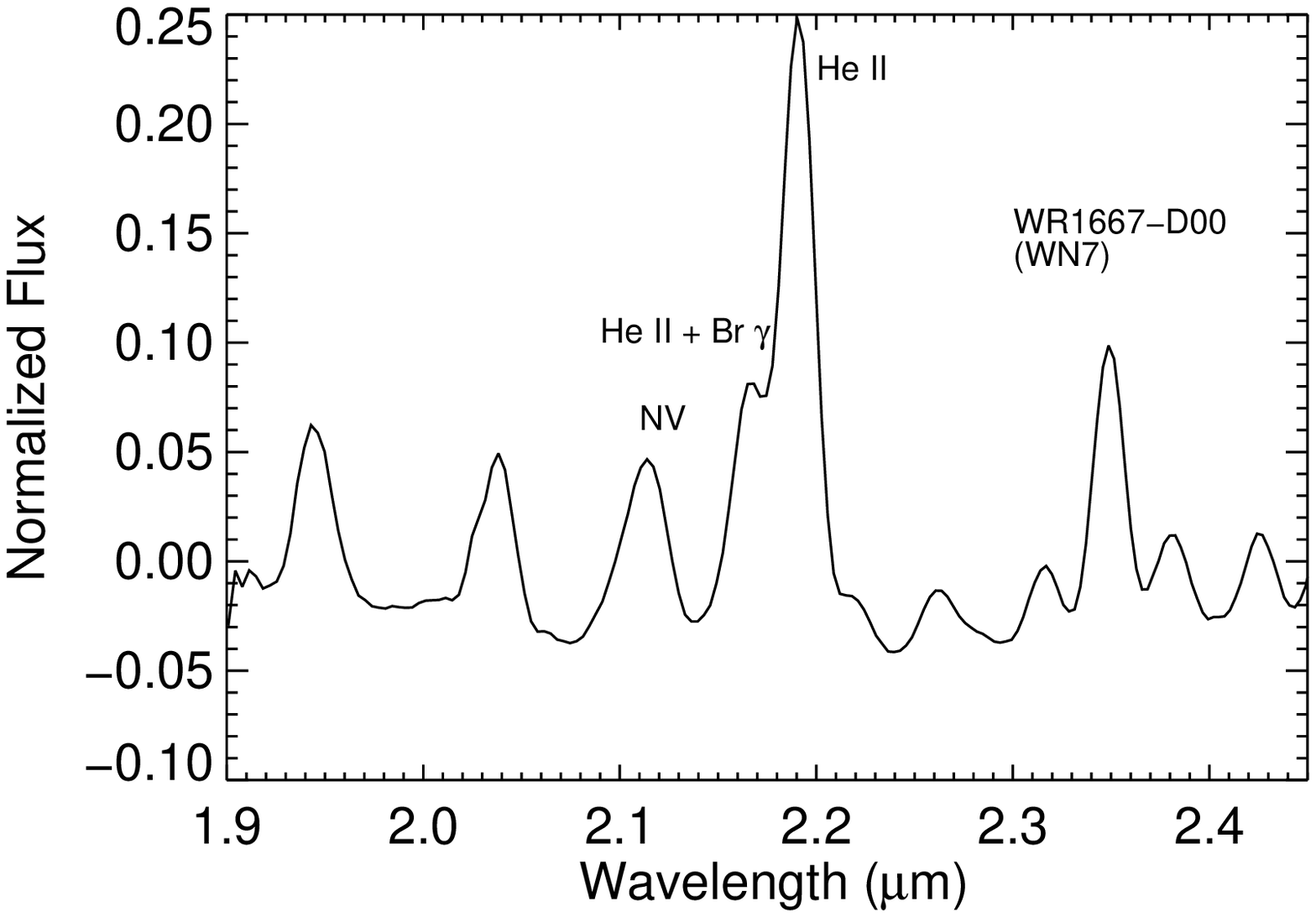,width=0.5\linewidth,clip=}
\end{tabular}
\caption{The four new WR stars found using the photometric criterion explained in the text. Details on each source are provided in Table~\ref{table:newWRs}. Prominent emission lines are noted. }
\label{fig:newWRs}
\end{figure*}

\begin{figure*}[!ht]
\begin{center}
\epsscale{1.2}
\includegraphics[width=.55\hsize]{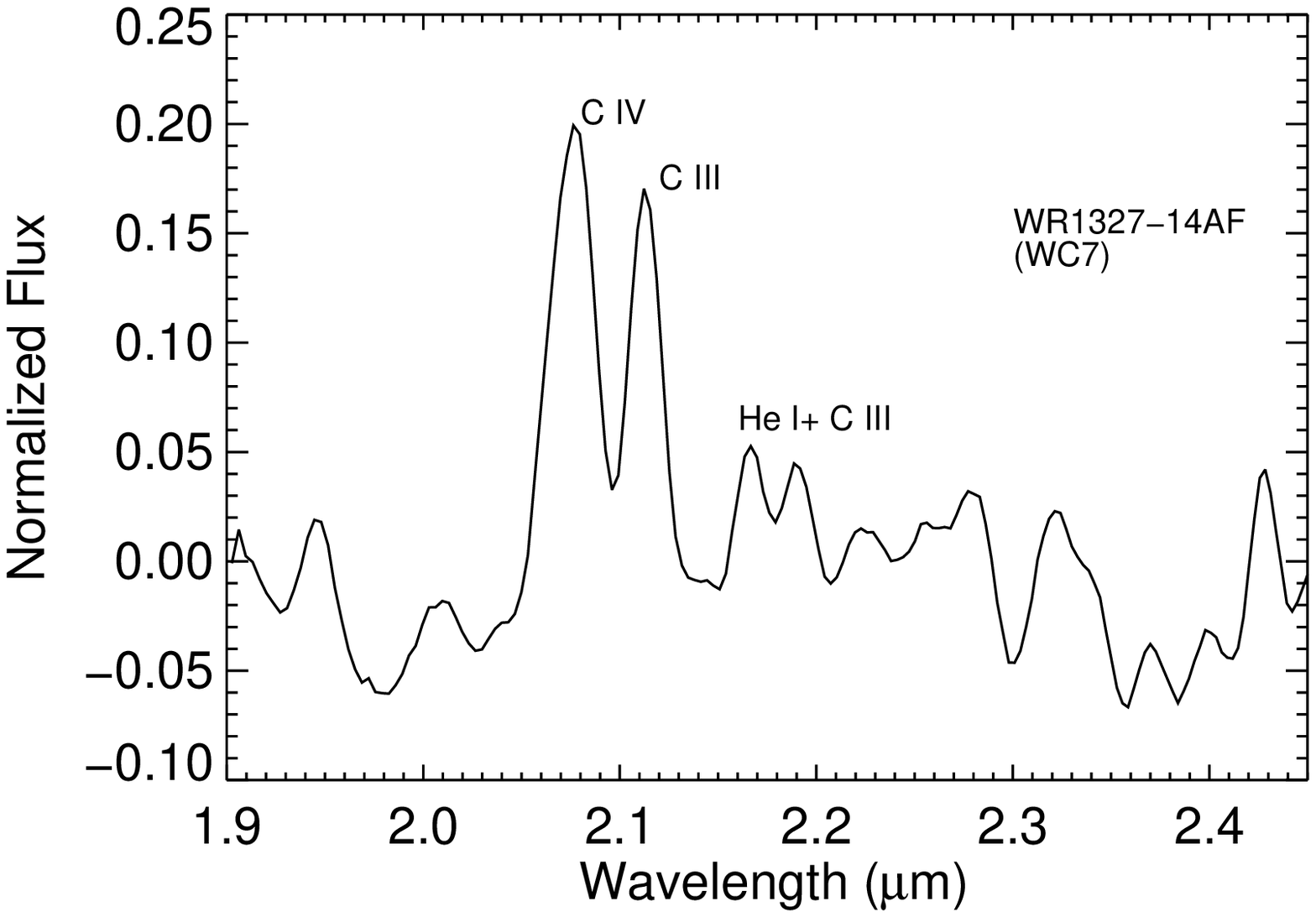}
\plottwo{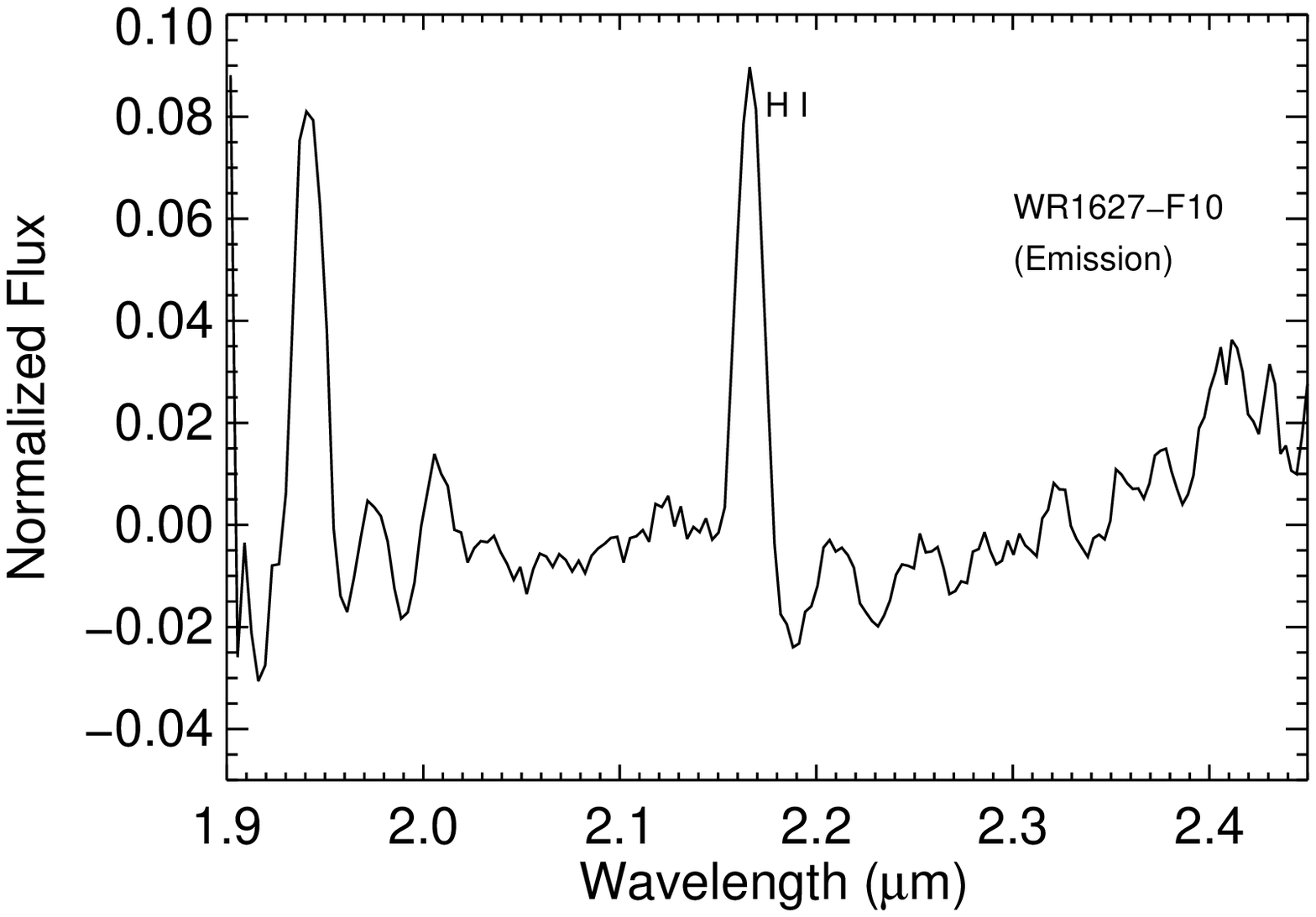}{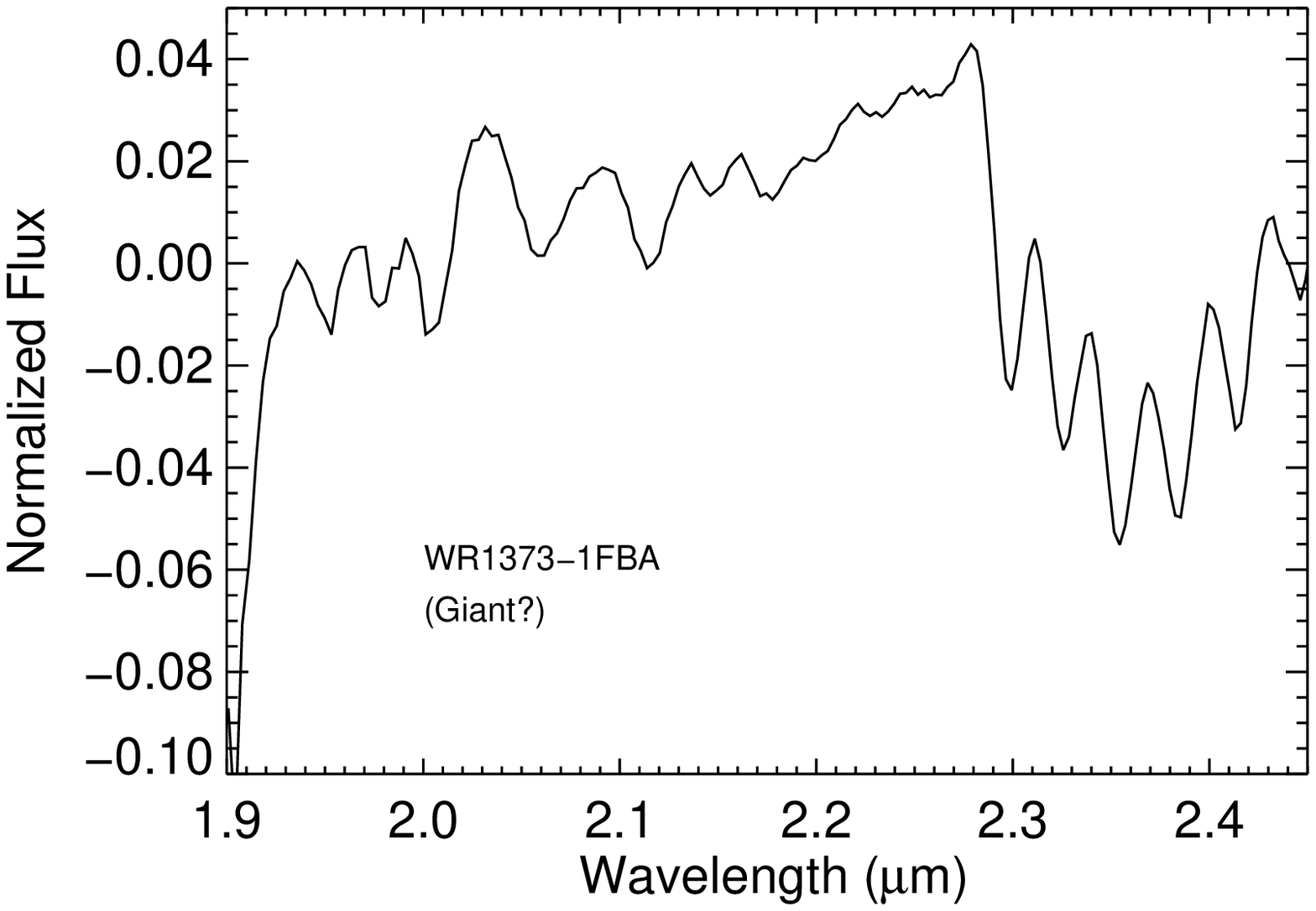}
\end{center}
\caption {{\bf Top:} A new WR star found just outside of the color color restrictions detailed in Table~\ref{table:colorspace} as it has a null K band detection.  {\bf Middle:} A new emission line source, and {\bf Bottom} A reddened background star.} 
\label{fig:duds}
\end{figure*}

Using SpeX time available through a collaborating program, we followed up on 6 strong candidates and found 5 to be WN  (4) or WC (1) stars with He I, He II, and/or Br$\gamma$ in emission, 1 to be an unidentified  emission line source (likely a Be star with H in emission), and 1 to be a hot giant star.  We also report one object, slightly outside the photometric criteria (with a ``null" $K_{s}$ detection), that also resulted in a new WR star. We show the $K_{s}$ spectra where emission line strengths are strongest in Figures~\ref{fig:newWRs} - \ref{fig:duds} as well as the position, photometry, and estimated spectral types in Table~\ref{table:newWRs}.  While the new sources demonstrate that the photometric constraints can identify strong candidates, there was one star which met all criteria but showed no sign of emission.  As discussed in previous works, this ``dud" demonstrates that there is no ``guaranteed" method to identifying WR -- or even emission line -- stars. 

\section{NIR AND MIR SPECTRAL FEATURES\label{spectra}}
WR stars have emission lines of CIV, HeI, HeII, and Br$\gamma$ that can reach $>$ 100 km s$^{-1}$ in equivalent width in the NIR.   \citet{Mauerhan11} investigated the impact of such strong emission lines on broad band NIR colors and found that they shifted WC stars by $\sim$0.3 mag in ($H$-$K_{s}$) and WN stars by $<$0.1 mag.  The effect is illustrated in the NIR color-color diagram of Figure~\ref{fig:nearir}  where WC stars are (on average) $>$0.5 mag above the field star loci, while WN stars stand out less, typically $>$0.1 mag.    

Moving out to longer MIR wavelengths, emission line strengths weaken, therefore this effect should diminish.   However, Figure~\ref{fig:nearmidIR} shows that WR ($W1$-$W2$) colors are consistently $\sim$ 0.3 mag redder than field or background sources.   The direction and extent of the reddening vector indicates that the spread in the X-direction (NIR color) can mainly be explained by interstellar reddening; but, the offset in the Y-direction (MIR excess) is unexpected.
    
\begin{figure*}
\centering
\begin{tabular}{cc}
\epsfig{file=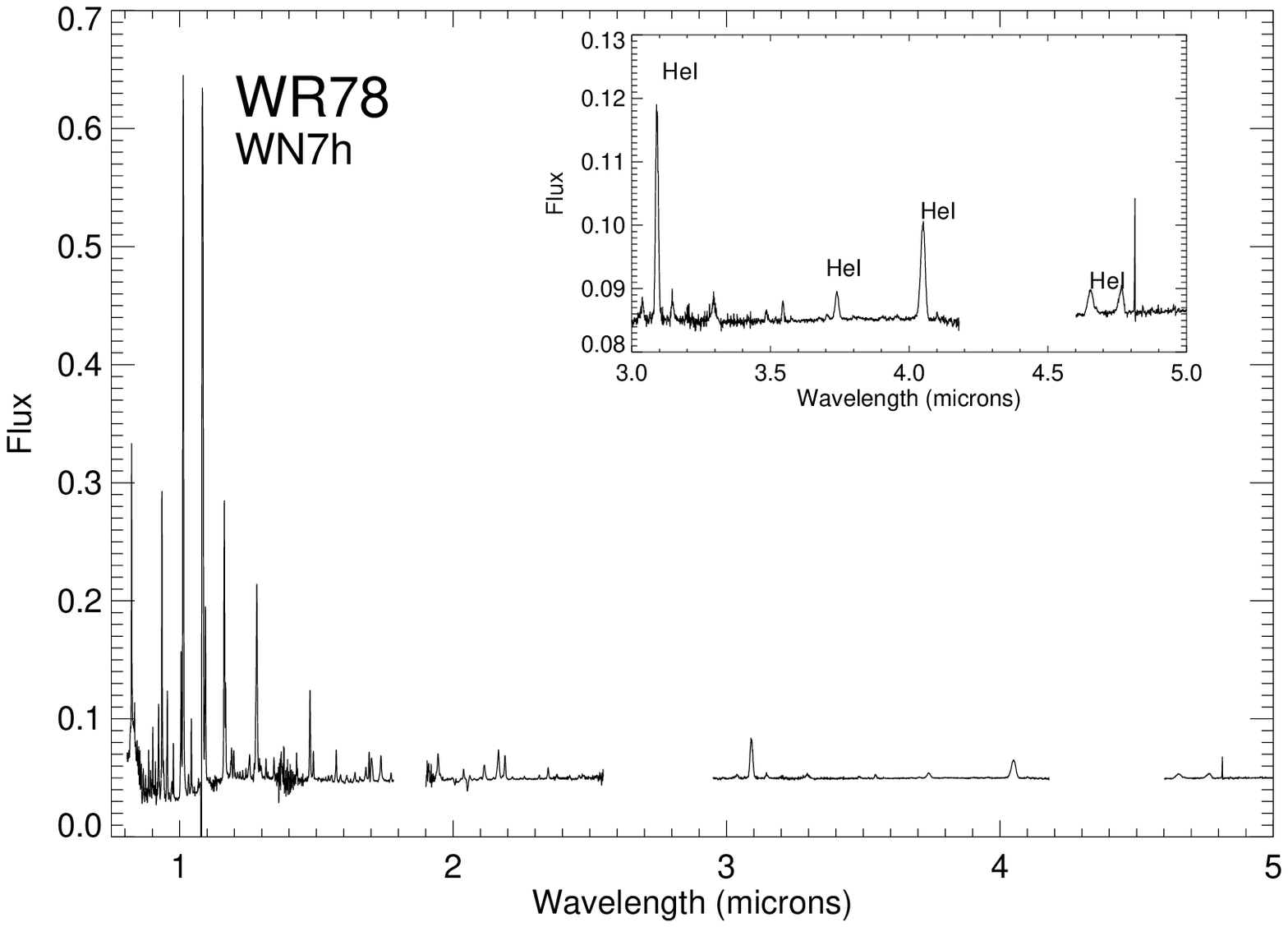,width=0.5\linewidth,clip=} &
\epsfig{file=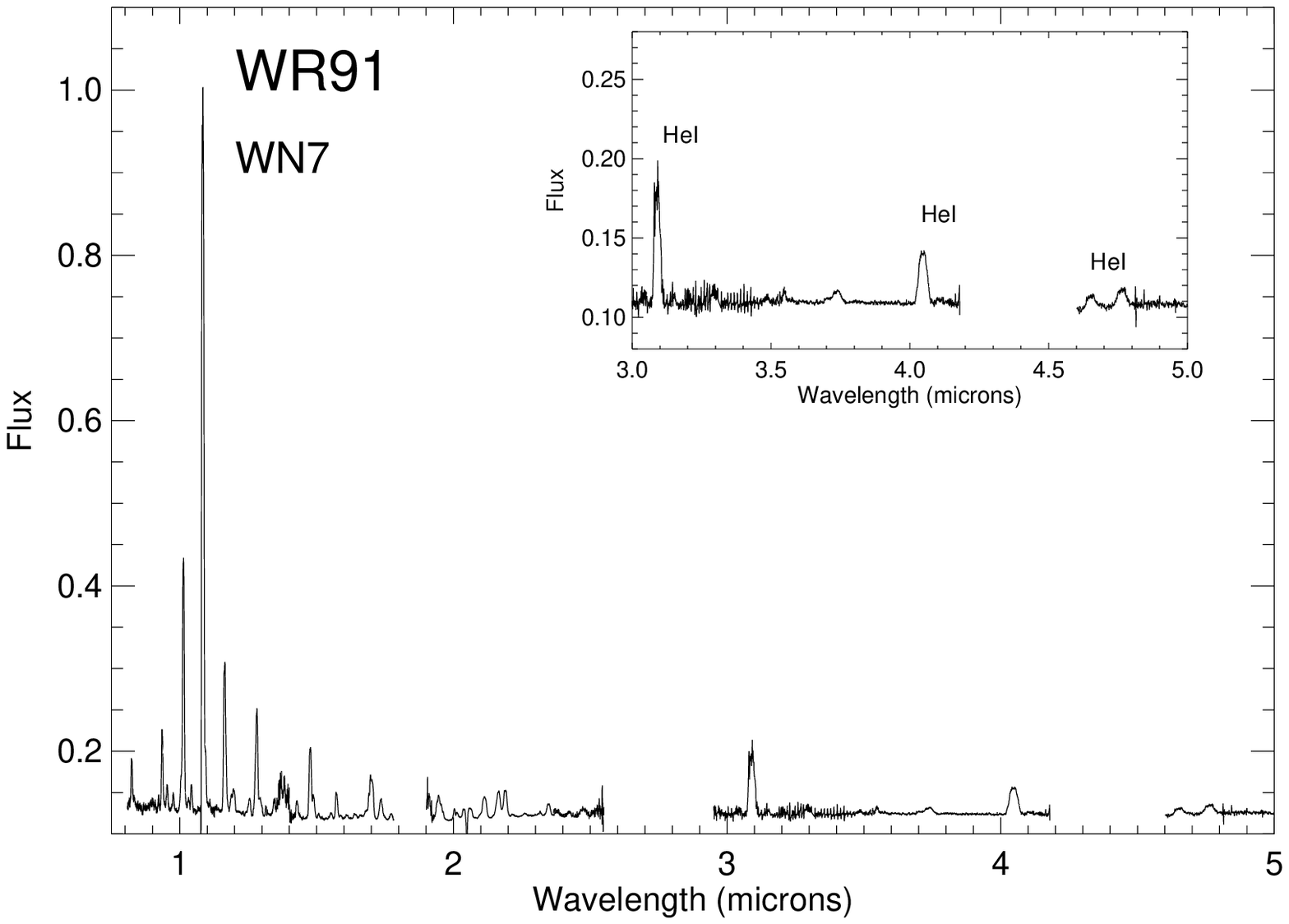,width=0.5\linewidth,clip=} \\
\epsfig{file=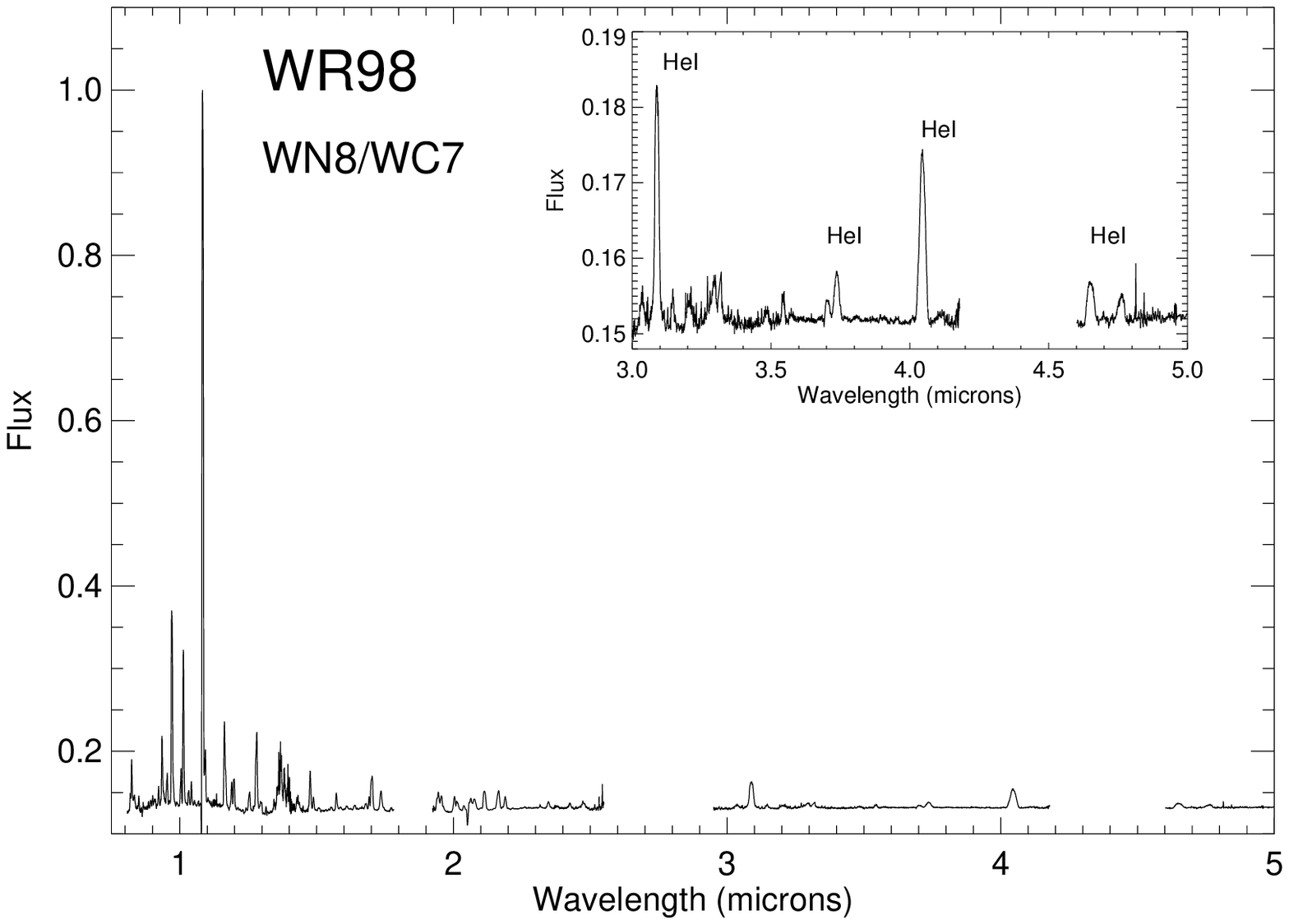,width=0.5\linewidth,clip=} &
\epsfig{file=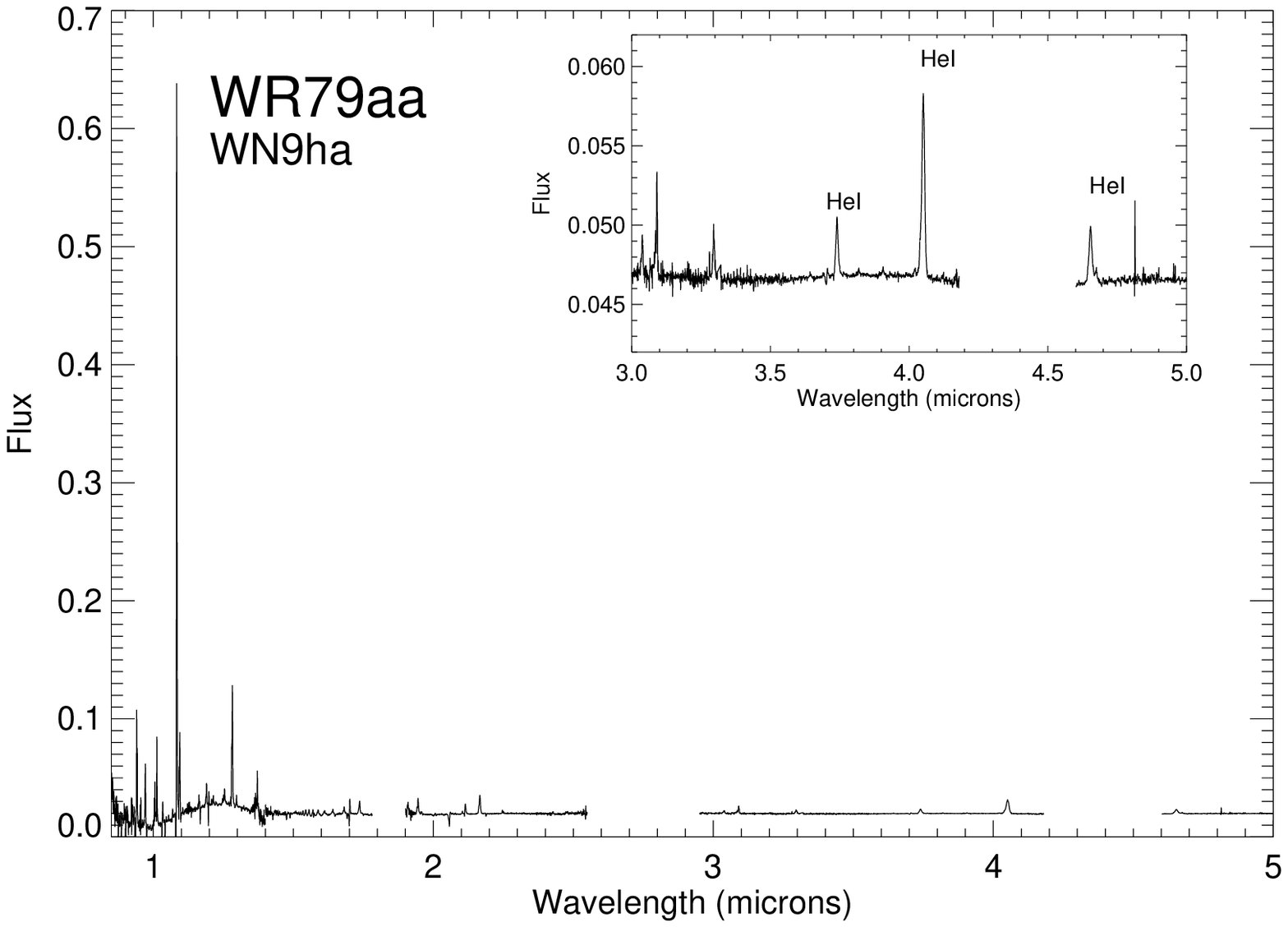,width=0.5\linewidth,clip=}
\end{tabular}
\caption{The 0.8 - 5.0$\mu$m spectra of four known WR stars. The 3.0-5.0$\mu$m portion is highlighted as an inset. These stars are highlighted on the NIR-MIR color-color diagram in Figure~\ref{fig:nearmidIR}. }
\label{fig:SXD1}
\end{figure*}

\begin{figure*}
\centering
\begin{tabular}{cc}
\epsfig{file=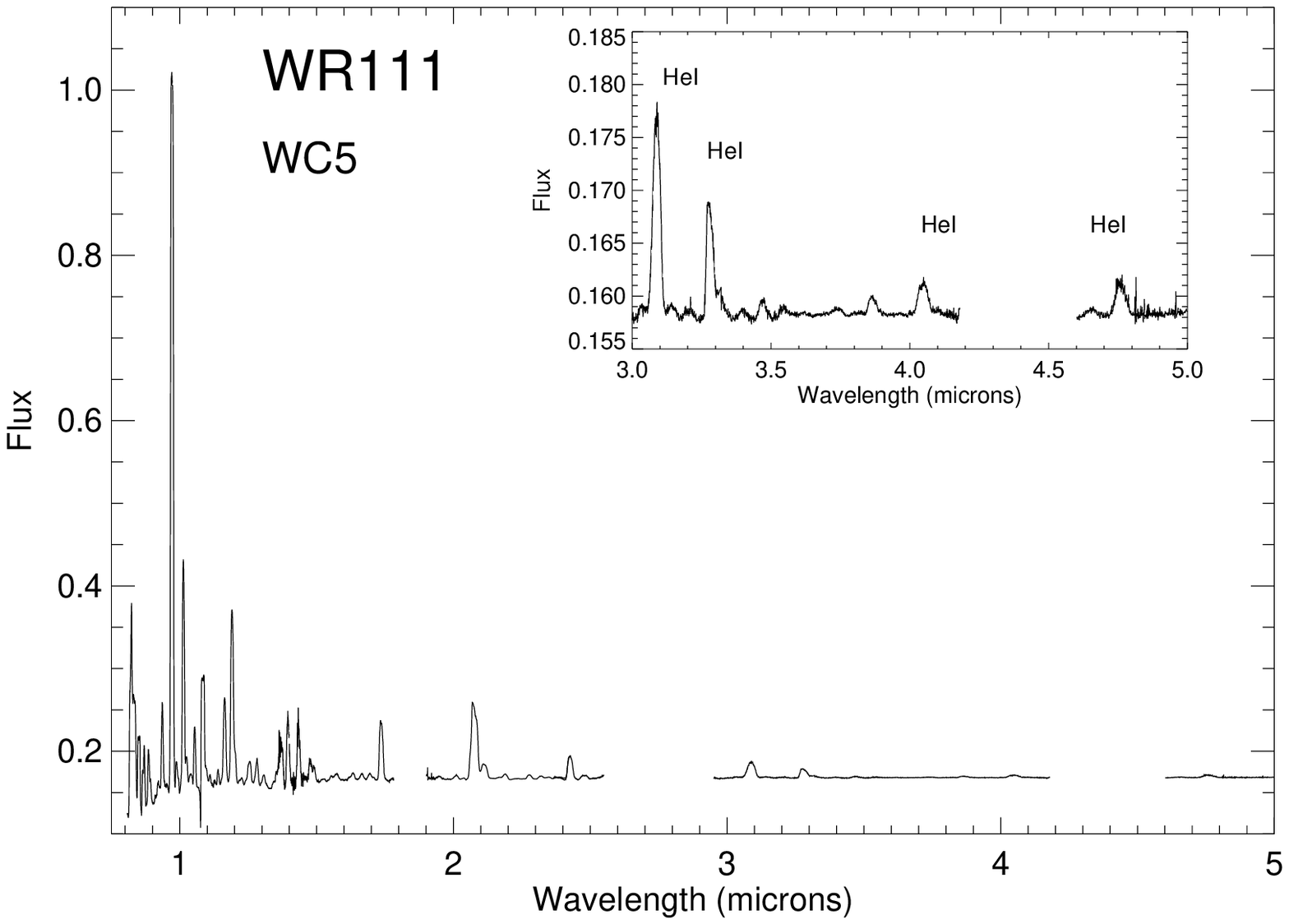,width=0.5\linewidth,clip=} &
\epsfig{file=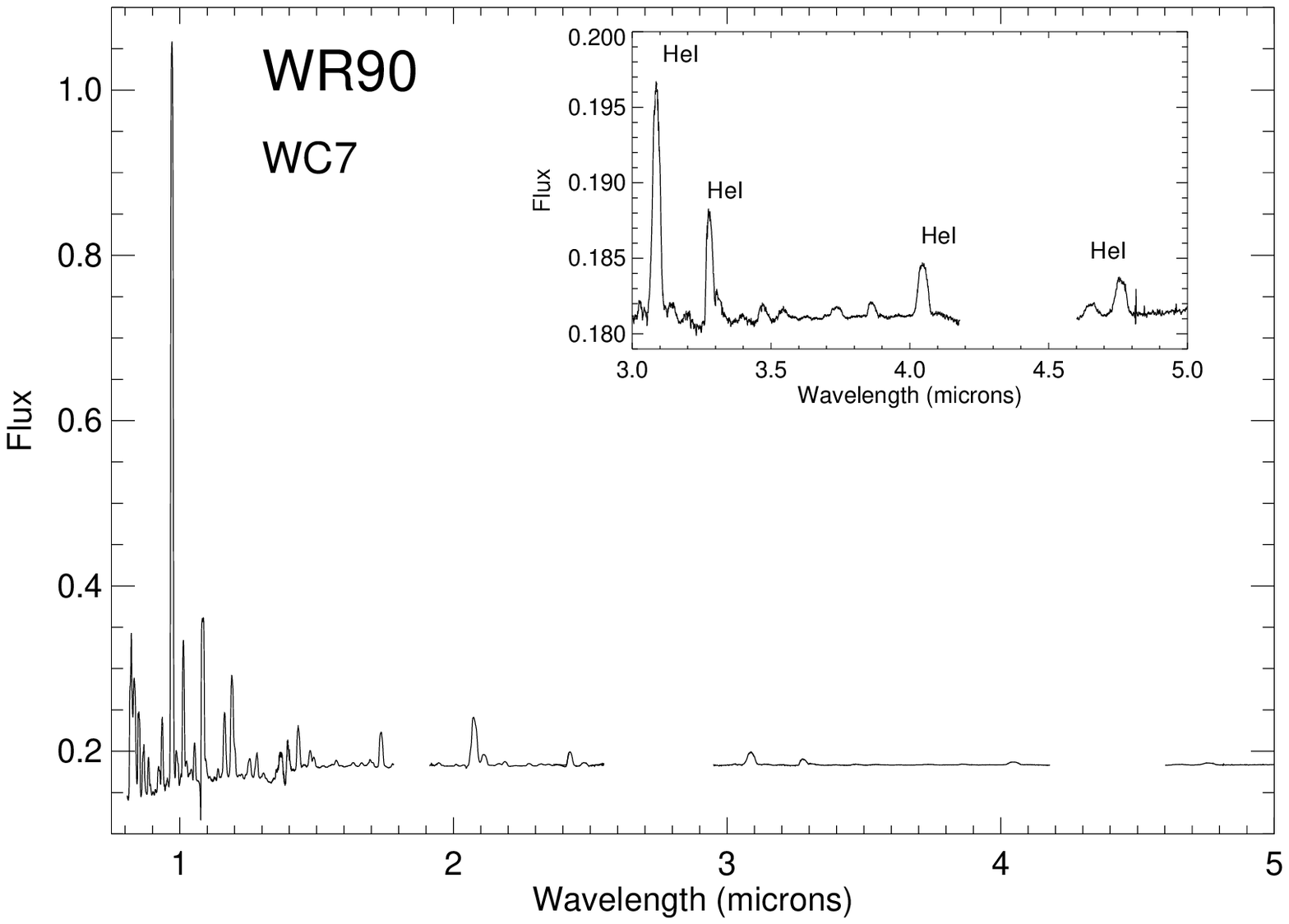,width=0.5\linewidth,clip=} \\
\epsfig{file=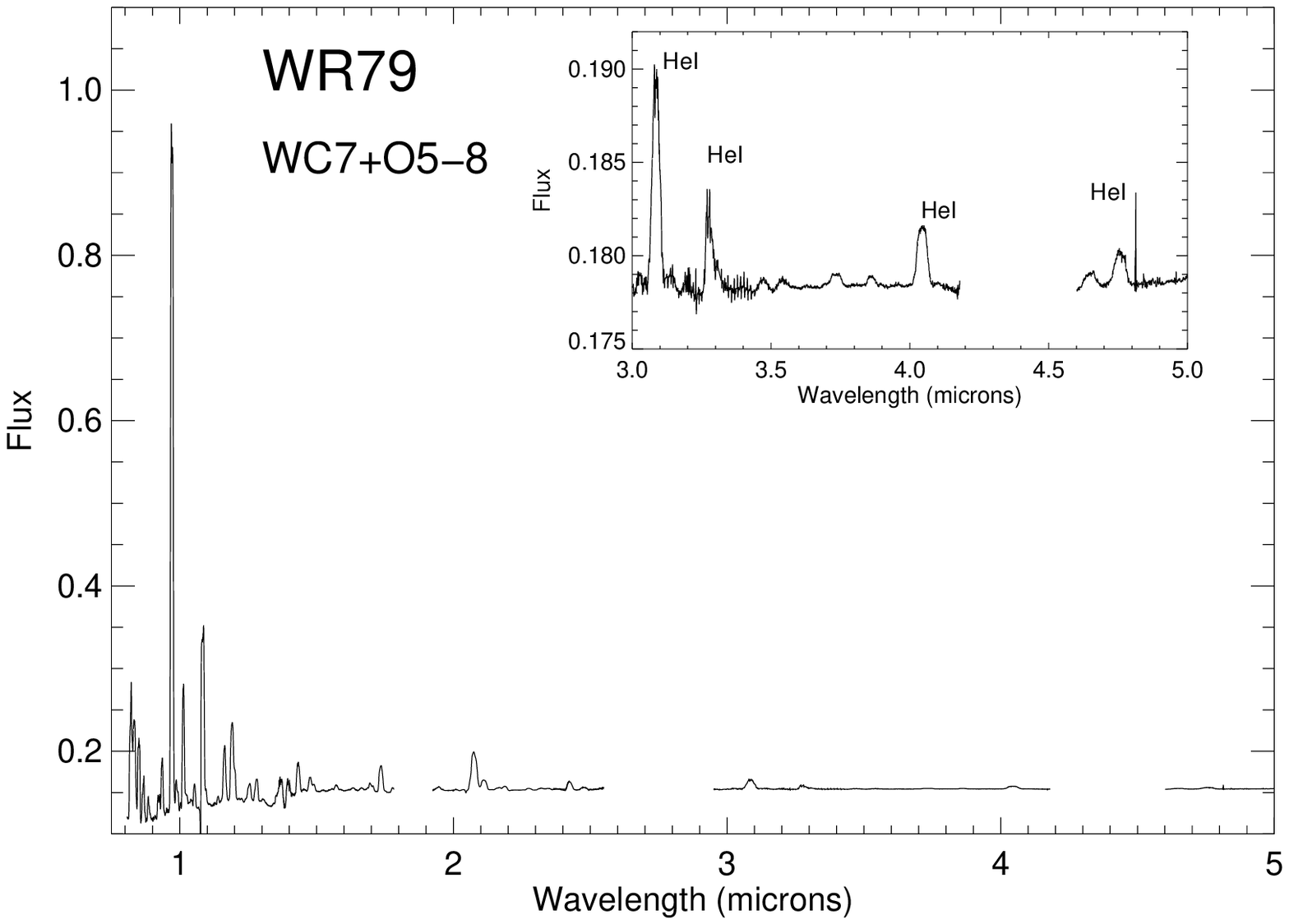,width=0.5\linewidth,clip=} &
\epsfig{file=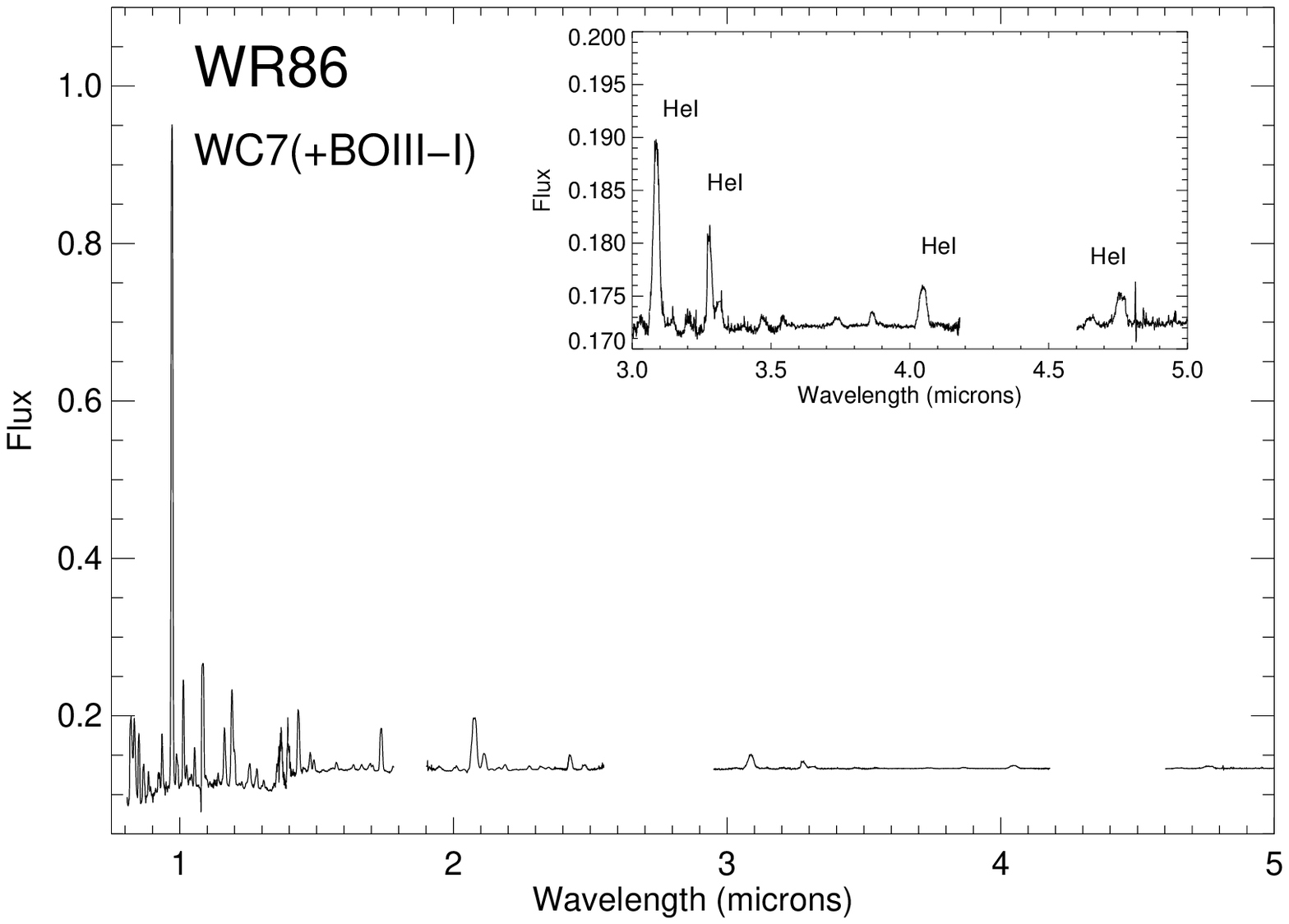,width=0.5\linewidth,clip=}
\end{tabular}
\caption{The 0.8 - 5.0$\mu$m spectra of four known WR stars. The 3.0-5.0$\mu$m portion is highlighted as an inset and their position on the NIR-MIR color-color is highlighted on Figure~\ref{fig:nearmidIR}. }
\label{fig:SXD2}
\end{figure*}

\begin{figure*}[!ht]
\begin{center}
\plottwo{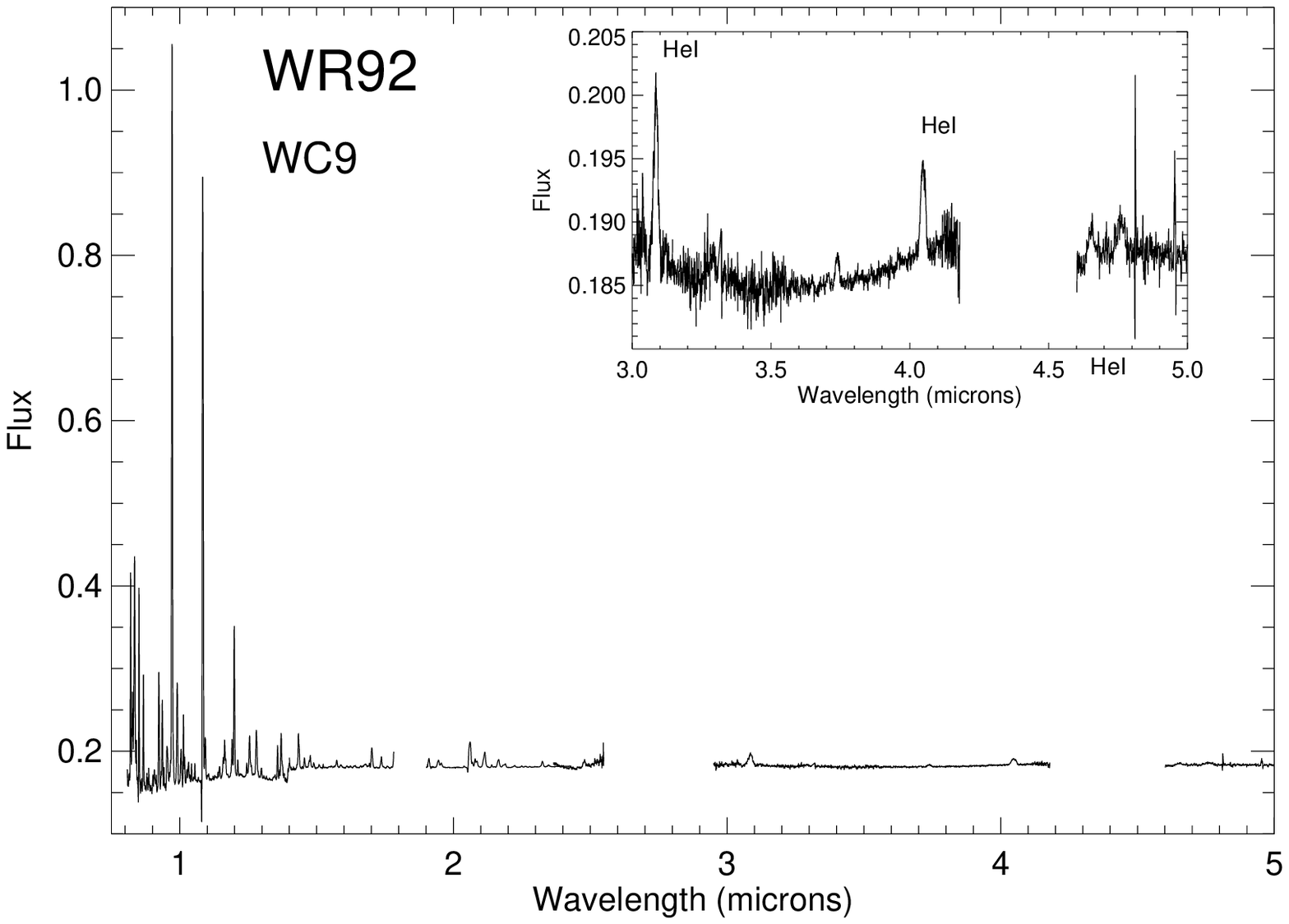}{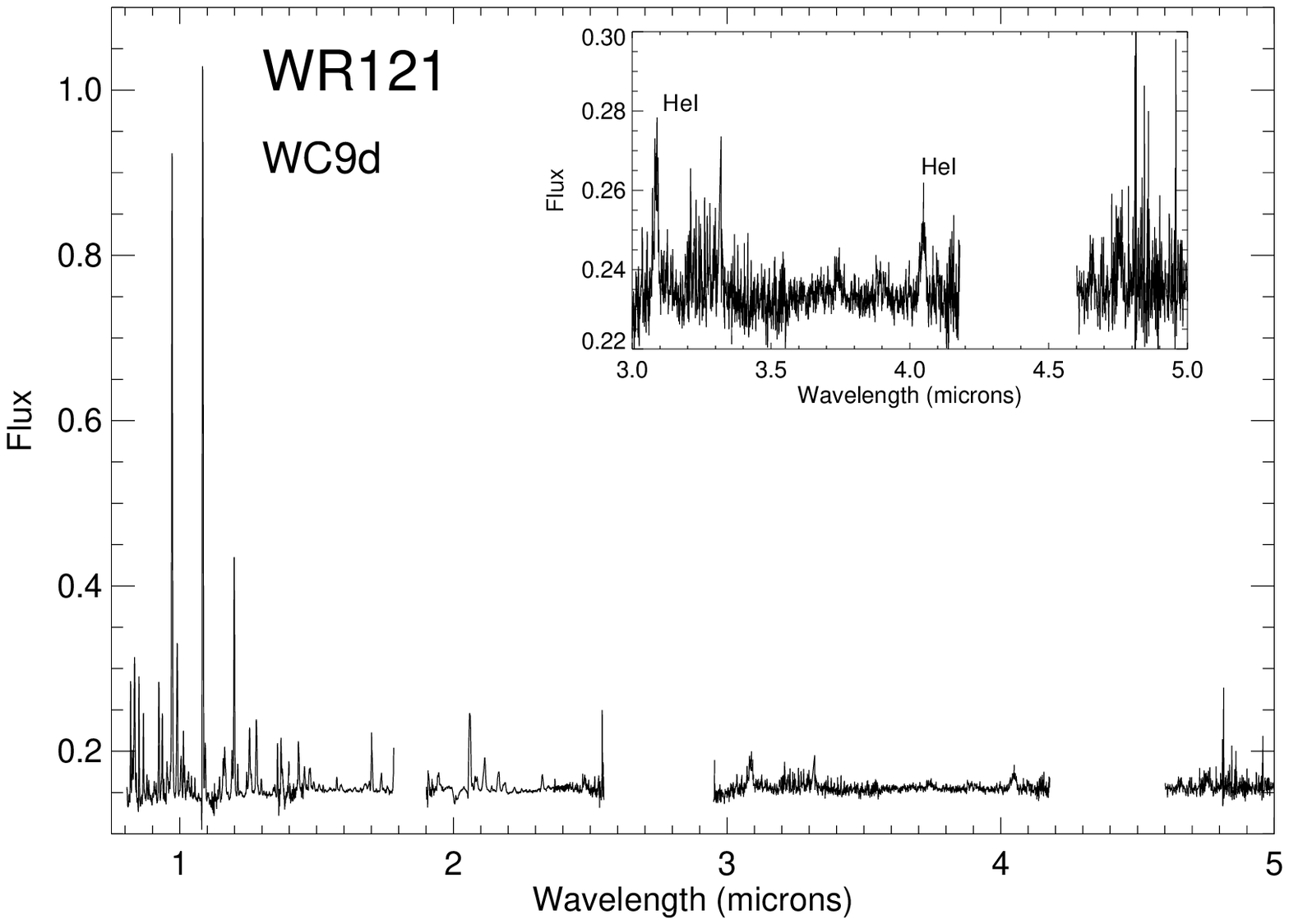}
\end{center}
\caption{The 0.8 - 5.0$\mu$m spectra of two known WR stars. The 3.0-5.0$\mu$m portion is highlighted as an inset. These stars are highlighted on the NIR-MIR color-color diagram in Figure~\ref{fig:nearmidIR}. }
\label{fig:SXD10}
\end{figure*}

To investigate the impact of emission lines on broadband colors, we used several hours of SpeX time to obtain both NIR and MIR spectra of 10 known, bright, and representative WR stars.  As shown in Table~\ref{SXD-LXD-Targets}, we focused on a sample that covered early to late WN and WC 's, one dusty source, and those that covered a range in ($W1-W2$) color.  We have highlighted the sources in Figure~\ref{fig:nearmidIR} as enlarged points and show the full  0.8-5.0 $\mu$m spectra in  Figures~\ref{fig:SXD1} - ~\ref{fig:SXD10}.  The spectra are normalized in the NIR between 0.8-1.4 $\mu$m.  As expected, the strongest emission lines from CIV, HeI, HeII, and Br$\gamma$ are seen shortward of 2$\mu$m.  Emission lines of primarily HeI persist into the mid-IR (3.0 -  5.0 $\mu$m  region overlapping with a portion of the $W1$ and $W2$ bands--highlighted in an inset for each object); however, as expected, their strength has greatly diminished.  

We examined the impact of these emission line features by integrating the flux over 3.0 - 5.0 $\mu$m and comparing results with and without prominent lines.  To evaluate the flux without lines, we linearly interpolated between the continuum around the prominent features at 3.0 and 4.0 $\mu$m.  Our analysis showed that the contribution from MIR emission lines is minimal ($<$0.01 mag), indicating that the excess seen in Figure~\ref{fig:nearmidIR} is related to the continuum.  As discussed in \citet{Mauerhan11}, such excess is linked to free-free electron scattering in dense, hot winds surrounding the central WR star. Moreover, many WC stars (especially of WC8 and later) show excess dust emission which (depending on the dust temperature) contaminates or ``enhances" the NIR to MIR color.  We suggest that a combination of free free emission and excess dust emission surrounding the WR stars leads to their position on Figure~\ref{fig:nearir}.   

\section{MEAN PHOTOMETRIC PROPERTIES\label{mean}}
\begin{figure*}[!ht]
\begin{center}
\epsscale{1.2}
\includegraphics[width=.55\hsize]{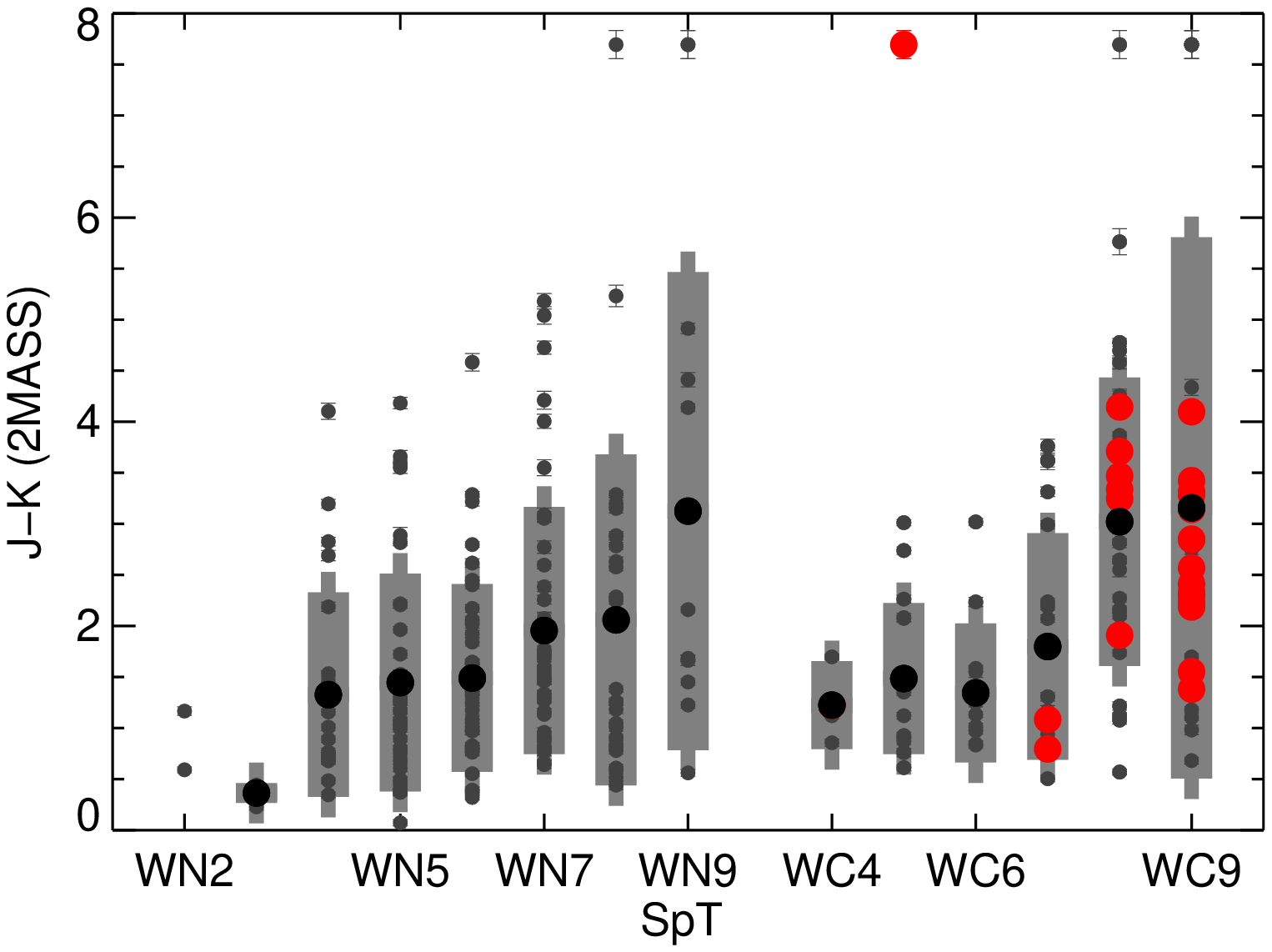}
\plottwo{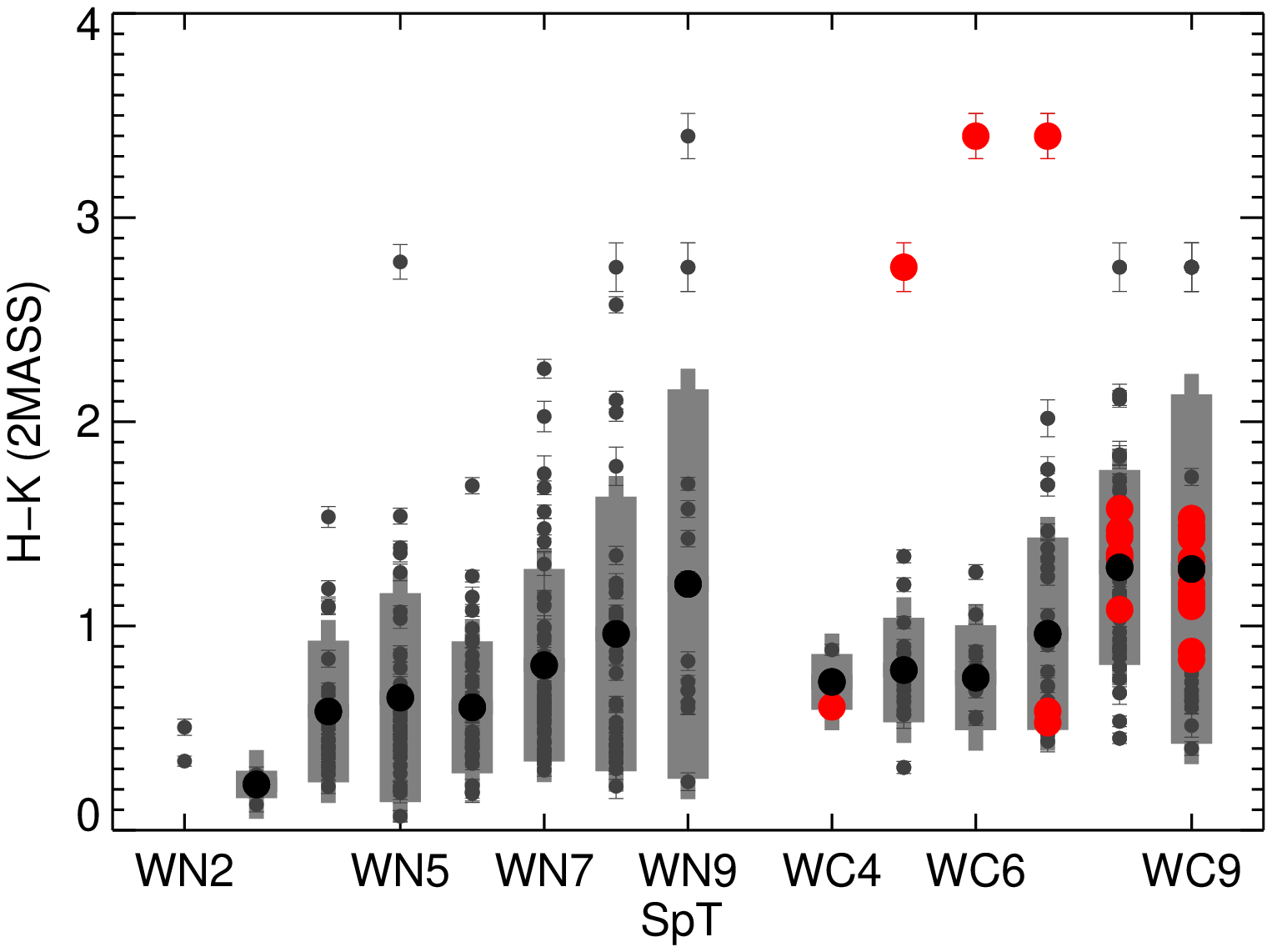}{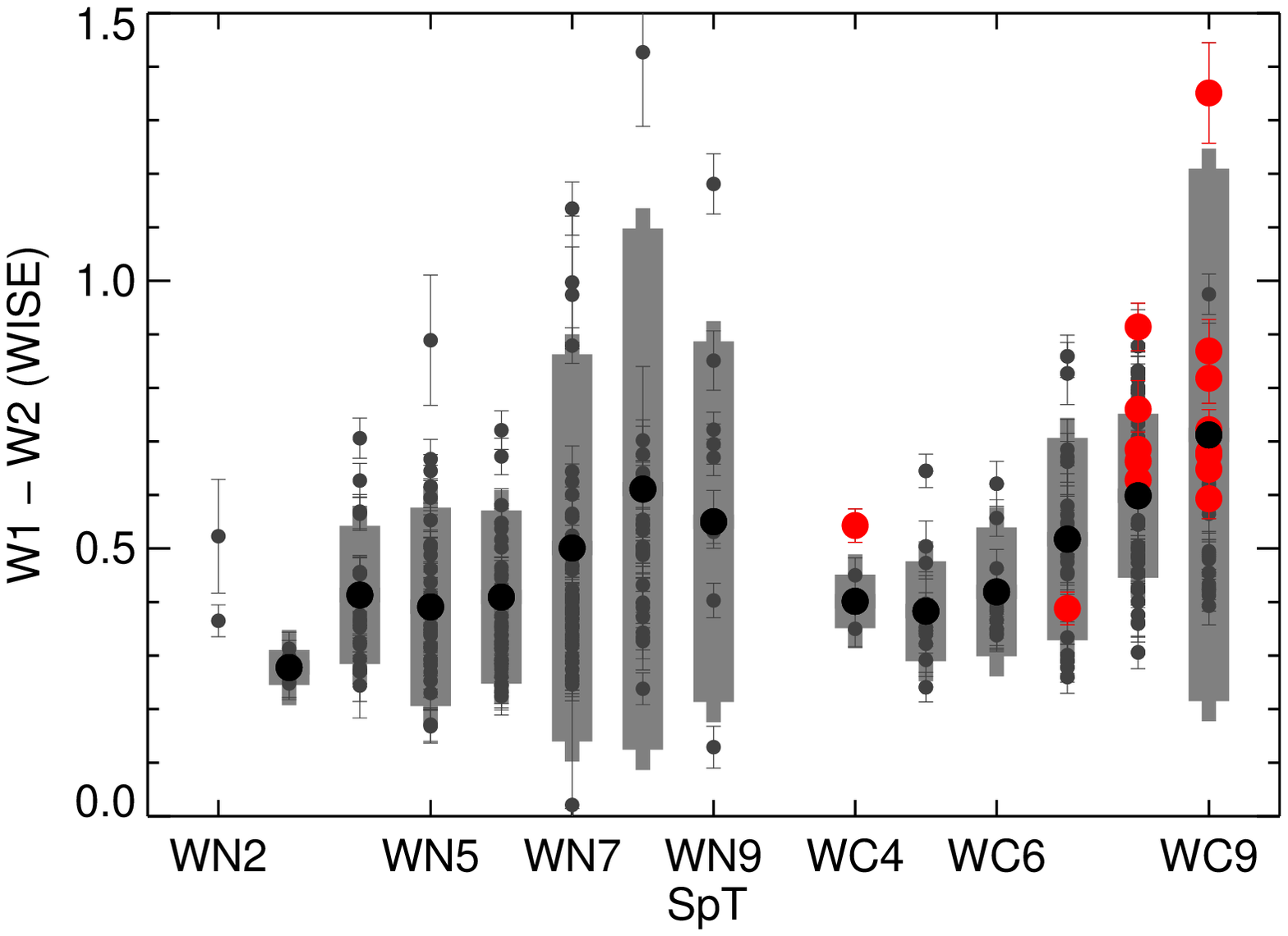}
\end{center}
\caption {The mean photometric properties of WN2 through WC9 stars are shown (note these have not been corrected for the reddening of each source).  Black dots at the center of each spectral subtype represent the mean color.  Grey bars in each demonstrate the standard deviation.  Red circles mark WC stars classified as dusty.  All quantities are reported in Table~\ref{meancolors}~-~\ref{meancolors2}.} 
\label{fig:NearIRSpt}
\end{figure*}

While having just a NIR (or just a MIR) broadband color alone is a poor indicator of a WR candidate (see discussion in \citealt{Hadfield07}), we investigated whether the difference in emission line strength between WC and WN stars could be discerned. In other words, if one could combine the NIR and MIR photometry (as discussed in this text) to confidently identify a WR candidate, can an individual NIR or MIR color then be used to differentiate the evolutionary stage of the source (i.e. WC from WN)?   As demonstrated in Figure~\ref{fig:nearir}, WC stars are redder than WN stars in ($H$-$K_{s}$) binned over ($J$-$H$).  To emphasize this, we highlight the regions that best identify the WC (upper) and WN (lower) populations on Figure~\ref{fig:nearir} and report the color space occupied by each in Table~\ref{table:colorspace}.  In Figure~\ref{fig:NearIRSpt} we present the NIR and MIR colors as well as averages and spreads for WC and WN stars as a function of spectral type. We use only WR stars that pass our photometric requirements in the given bands (see Table~\ref{table:addconstraints}) and are not classified as ``Dusty'' (although dusty sources are over plotted to demonstrate their positions).  Quantitative information is  presented in Tables~\ref{meancolors}-\ref{meancolors2} although we note that these have not been corrected for the reddening of each source.   As one moves through the subtypes of WN stars as well as WC stars, from early subtypes to later, the mean NIR and MIR color moves redward.  However, there is significant overlap between each phase, therefore little information can be drawn from an individual color. The increased reddening in both the ($J$-$K_{s}$) and ($W1$-$W2$) from early WR subtypes to late as well as the fact that the reddest (non-dusty) individual sources in each color are the latest spectral types, suggests that current NIR and MIR broadband searches are mildly biased toward finding late-type WR (either WN or WC) stars at greater distances.

\section{SUMMARY\label{conclusions}}
We have cross-correlated a list of known WR stars, PN, CV, and Be stars with the 2MASS and WISE online photometric catalos to investigate the NIR and MIR characteristics of emission line objects.  Investigating color-color plots by using a combination of $J,H,K_{s}, W1,W2, W3$, and $W4$ photometry, we present the most prominent diagnostic colors that distinguish emission line objects from the field stellar population as well as from each other.  We find that emission line objects, and WR stars in particular, are well distinguished from the background stellar population in the ($W1$-$W2$) vs. ($J-K_{s}$) color-color diagram.  PN can be easily distinguished from other emission line objects by their excessive emission in ($W1$-$W3$).  We suggest that identifying PN photometric candidates could be accomplished by searches for objects with ($W1$-$W3$)  $>$ 4.0 mag.  

WR stars can be further distinguished from emission line objects (PN, Be and CV stars) by including restrictions in (1)  $(K_{s}-W3)$ vs. $(J-K_{s})$, (2) ($J$-$H$) vs. ($H$-$K_{s}$), and (3) ($W1$-$W3$) vs ($J$-$K_{s}$) color color diagrams.  Supplementing such plots with narrow-band photometry (see \citealt{Shara09,Shara12}), is an extremely powerful tool for detecting new WR stars.  As a proof of concept, we followed up on 6 sources and detected and characterized 4 new WN stars and 1 new WC star.  We note that \citet{Mauerhan11} found that Be stars may continue to contaminate even with strict restrictions in the NIR and MIR.  While there are now significant numbers of WR stars known in the Galactic plane, the same cannot be said for equally interesting emission line samples. In order to fully assess the diagnostic plots presented here, a more extensive sample of distant (hence reddened) emission line objects must be detected and included.  

Mean photometric properties of known WR stars are presented in the ($J-K_{s}$), ($H-K_{s}$), and ($W1$-$W2$) colors (although without corrections for individual source reddening).  The progression toward redder ($J-K_{s}$) and ($W1$-$W2$) color for early WC or WN to late WC or WN indicates that current studies have been mildly biased toward detecting more late-type WR stars at greater distances.  

Both NIR and MIR spectra were presented for a select sample of 10 known WR stars.  Strong emission lines of HeI, HeII, CIV, Br$\gamma$ are seen throughout the 0.8-5.0 $\mu$m range, albeit with strengths that diminish with increasing wavelength.  We find the strength of emission in the MIR cannot account for the excess above the background field sample seen in the ($J-K_{s}$) versus ($W1-W2$) color-color diagram.  In agreement with the result proposed by \citet{Mauerhan11}, we find that the excess is primarily due to continuum contributions enhanced by free-free electron scattering in the surrounding environment.

\acknowledgments{We acknowledge the generous support of Hilary and Ethel Lipsitz, staunch friends of the Astrophysics Department of AMNH.  J. Faherty was supported by NSF IRFP grant 0965192.  J. Faherty would like to thank J. Mauerhan for a useful discussion on Wolf Rayet stars.  We also thank the anonymous referee for a thorough and helpful report.  This publication has made use of the data products from the Two Micron All-Sky Survey, which is a joint project of the University of Massachusetts and the Infrared Processing and Analysis Center/California Institute of Technology, funded by the National Aeronautics and Space Administration and the National Science Foundation. This research has made use of the NASA/ IPAC Infrared Science Archive, which is operated by the Jet Propulsion Laboratory, California Institute of Technology, under contract with the National Aeronautics and Space Administration. We acknowledge receipt of observing time through NASA IRTF.}

\bibliographystyle{apj}
\bibliography{paper}

\clearpage

\clearpage

\clearpage

\begin{deluxetable*}{lc}
\label{tab:tab1}
\tabletypesize{\scriptsize}
\tablecaption{Photometric Requirements for Candidate WR stars\label{table:addconstraints}}
\tablewidth{0pt}
\tablehead{
\colhead{Catalog} &
\colhead{Constraint} \\
\colhead{(1)} &
\colhead{(2)}  \\
}

\startdata
2MASS & $J$, $H$, and $K_{s}$ band detection (uncertainty not "null``)\\
2MASS & $J$ $>$ 5.5, $H$ $>$ 5.0, $K_{s}$ $>$ 4.5 (to avoid saturated sources)\\
WISE & $W1,W2$, and $W3$ band detection (uncertainty not "null``)\\
WISE & $W1$ $>$ 2.0, $W2$ $>$ 1.5, $W3$ $>$ -3.0 (to avoid saturated sources)\\
WISE & SNR on $W1,W2,W3$ $>$ 3 and SNRI\tablenotemark{a} $>$ 7\\
\enddata
\tablenotetext{a}{The SNR of individual detections in any band must be $>$ 3 for at least 7 images}
\end{deluxetable*}

\begin{deluxetable*}{lcccrrrrrrrrrrr}
\label{tab:tab1}
\tabletypesize{\scriptsize}
\tablecaption{ Details of SpeX Observations\label{SpeX}}
\tablewidth{0pt}
\tablehead{
\colhead{Name} &
\colhead{SpT} &
\colhead{Date} &
\colhead{Mode}&
\colhead{Int Time}&
\colhead{Coadds}&
\colhead{Images}&
\colhead{Airmass} \\
 &
 &
 &
 &
\colhead{(s)} &
&
 &
 \\
\colhead{(1)} &
\colhead{(2)} &
\colhead{(3)} &
\colhead{(4)} &
\colhead{(5)} &
\colhead{(6)} &
\colhead{(7)} &
\colhead{(8)} \\
}
\startdata
WR78	&	WN7h 		&14 July 2011		&SXD	& 10 & 1	& 6 &2.50\\
	&	 		&13 July 2011		&LXD	&	0.5&10&4&2.33\\
	\hline
WR91	&	WN7  		&14 July 2011		&SXD	& 20  & 1    & 4 & 2.05\\
	&	  		&13 July 2011		&LXD	&0.5&30&16&2.01\\
\hline
WR98	&	WN8/WC7	&14 July 2011		&SXD	&30	&1	&	4	&1.76\\
&		&13 July 2011		&LXD&0.5&30&8&1.70\\
\hline
WR79a 	&	WN9ha 		&14 July 2011		&SXD	&5	&1	&	6	&2.37\\
	&	 		&13 July 2011		&LXD&0.5&20&8&2.23\\
\hline
WR111	&	WC5  		&13 August 2009	&SXD	&60&1&4&1.34\\
	&	  		&13 July 2011		&LXD&0.5&30&8&1.40\\
\hline
WR90	&	WC7			&14 July 2011		&SXD	&10&1&4&2.52\\
	&				&13 July 2011		&LXD	&0.5&30&8&2.58\\
\hline
WR79	&	WC7+O5-8 	&14 July 2011		&SXD	&5&1&6&2.44\\
&	 	&13 July 2011		&LXD &0.5&20&8&2.32\\
\hline
WR86	&	WC7 (+B0III-I)  &14 July 2011		&SXD	&30&1&4&2.22\\
	&	  &13 July 2011		&LXD&0.5&30&8&2.59\\
\hline
WR92	&	WC9 		&14 July 2011		&SXD	&30&1&4&2.34\\
	&			&13 July 2011		&LXD&0.5&30&16&2.32\\
\hline
WR121	&	WC9d  		& 27 August 2009	&SXD	&30&1&4&1.50\\
	&	  		&13 July 2011		&LXD&0.5&30&8&1.12\\
	\hline
WR1343-193E & WN6 & 18 August 2012 & LowRes15 & 30 & 1 & 2 & 2.20\\
WR1583-B73 & WN6 & 18 August 2012  & LowRes15 & 30 & 1 & 2 & 2.22\\
 WR1667-D00 & WN7 & 30 August 2012  & LowRes15 & 90 & 1 & 2 & 1.59\\
WR1361-1583 & WN9 & 31 August 2012  & LowRes15 & 90 & 1 & 2 & 1.77\\
WR1327-14AF & WC7 & 30 August 2012  & LowRes15 & 90 & 1 & 2 & 1.17\\
WR1627-F10 & Emmision & 01 September 2012  & LowRes15 & 90 & 1 & 2 & 1.42\\
WR1373-1FBA & Giant? & 31 August 2012  & LowRes15 & 70 & 1 & 2 & 1.14\\

\hline
\hline
\enddata
\end{deluxetable*}

\begin{deluxetable*}{ccccrrrrrrrrrrr}
\label{tab:tab1}
\tabletypesize{\scriptsize}
\tablecaption{ Color Space for Identifying WR stars\label{table:colorspace}}
\tablewidth{0pt}
\tablehead{
\colhead{Priority Rank}&
\colhead{Parameters} &
\colhead{Notes} \\
\colhead{(1)} &
\colhead{(2)} &
\colhead{(3)} \\}
\startdata
1 &       ($W1-W2$)    $>$      0.09 x  ($J-K_{s}$)  +  0.03& Highlights emission line (alone eliminates 95\% of Bkgrnd)\\
         \hline
2    &     ($H-K_{s}$)         $>$      0.50 x  ($J-H$) +  0.02&Highlights emission line (w/ below eliminates 92\% of Bkgrnd)\\
     &    ($H-K_{s}$)         $<$      0.60 x  ($J-H$) + 0.63&\\
         \hline
3&         Q1             $<$      0.1& Highlights emission line (w/ below eliminates 98\% of Bkgrnd)\\
    &     Q2             $<$     -1.0&\\
     &    Q2             $>$     11.25 x Q1 - 2.38&\\ 
\hline

4 &        ($W1-W3$)    $<$      0.78 x  ($J-K_{s}$ ) +  1.90&Eliminates PN\\
         \hline
5 &         ($J-K_{s}$)          $<$      2.47 x  ($K_{s} -W3$) - 1.29& W/ below eliminates 36\% of CV and 73\% of Be stars\\
    &           ($J-K_{s}$)          $>$      0.30&\\
\hline

 &        ($H-K_{s}$)         $>$      0.50 x  ($J-H$) +  0.02&Isolates 90\% of WN stars\\
 &        ($H-K_{s}$)         $<$      0.50 x  ($J-H$) + 0.27&\\
\hline
   &      ($H-K_{s}$)         $>$      0.50 x  ($J-H$) +  0.27&Isolates 80\% of WC stars\\
 &        ($H-K_{s}$)         $<$      0.60 x  ($J-H$) + 0.63&\\
\enddata
\end{deluxetable*}

\clearpage
\begin{landscape}
\begin{deluxetable*}{lcccrrrrrrrrrrr}
\label{tab:tab1}
\tabletypesize{\scriptsize}
\tablecaption{ Sample of WR stars with Shells\label{Table:shells}}
\tablewidth{0pt}
\tablehead{
\colhead{Name} &
\colhead{RA} &
\colhead{DEC} &
\colhead{SpT} &
\colhead{Shell at $W3$} &
\colhead{Shell at $W4$} &
\colhead{Comments}& 
\colhead{Reference}\\
\colhead{(1)} &
\colhead{(2)} &
\colhead{(3)} &
\colhead{(4)} &
\colhead{(5)} &
\colhead{(6)} &
\colhead{(7)} &
\colhead{(8)} \\
}
\startdata
WR7			&07 18 29.13&-13 13 01.5&WN4 & y&y&Two layer, full circular shell&1\\    
WR8			&07 44 58.22&-31 54 29.6&WN7/WCE+?&y&y&Diffuse shell in $W3$, circular shell in $W4$&1\\
WR35		&11 00 22.10&-61 13 51.0&WN6h+OB?&n&y&Three quarter shell with bright western side in $W4$&1\\ 
--			&15 35 26.52&-56 04 12.3&WN7&n&y&Faint circular shell in $W4$&2\\
1093-53		&16 32 12.98&-47 50 35.8&WN5b&n&y&Three quarter bubble in $W4$&3\\
1093-140LB	&16 32 48.99&-47 44 53.2\tablenotemark{a}&WN9&y&y&Three quarter, circular shell in $W3$ faint in $W4$&4\\
WR75ab		&16 33 48.74&-49 28 43.5&WN7h&y&y&Bright circular shell in $W3$ and $W4$&1\\
---			&18 42 08.27&-03 51 02.9&WN8-9h&n&y&Circular bubble with bright lobe in $W4$&5\\
\hline
WR16                 &09 54 52.91&-57 43 38.3&WN8h&y&y& Point source and large Shell in $W3$ and $W4$&1\\
WR40                 &11 06 17.21&	-65 30 35.2&WN8h&y&y&  Circular shell/bubble in $W4$&1\\	
\hline
\hline
\enddata
\tablenotetext{a}{The coordinates reported in \citet{Shara12} (and subsequently in the SpeX image acquired for that target) are as noted above.  However, based on the WISE shell, we believe this source is actually 2MASSJ16324851-4745062.}
\tablecomments{References: 1=\citet{van-der-Hucht06}, 2=\citet{Mauerhan09}, 3=\citet{Shara09}, 4=\citet{Shara12}, 5=\citet{Mauerhan10a}}
\end{deluxetable*}
\clearpage
\end{landscape}

\clearpage
\begin{landscape}
\begin{deluxetable*}{lccccccccccccrr}
\label{tab:tab1}
\tabletypesize{\tiny}
\tablecaption{Properties of New WR stars \label{table:newWRs}}
\tablewidth{0pt}
\tablehead{
\colhead{Name} &
\colhead{RA} &
\colhead{DEC} &
\colhead{J} &
\colhead{H} &
\colhead{K} &
\colhead{W1} &
\colhead{W2} &
\colhead{W3} &
\colhead{W4} &
\colhead{SpT} \\
&
&
&
\colhead{2MASS} &
\colhead{2MASS} &
\colhead{2MASS} &
\colhead{WISE} &
\colhead{WISE} &
\colhead{WISE} &
\colhead{WISE} \\
\colhead{(1)} &
\colhead{(2)} &
\colhead{(3)} &
\colhead{(4)} &
\colhead{(5)} &
\colhead{(6)} &
\colhead{(7)} &
\colhead{(8)} &
\colhead{(9)} &
\colhead{(10)} &
\colhead{(11)} \\
}
\startdata
WR1343-193E&18 02 46.23 &-22 36 39.7&14.926$\pm$       0.065&    12.690$\pm$        0.054&    11.447$\pm$        0.046& 10.427$\pm$     0.041&   9.854$\pm$     0.035&  10.004$\pm$      null&  7.250$\pm$      null&WN6\\
WR1583-B73& 19 00 05.09& +03 47 27.1&16.781$\pm$        0.158   & 13.606$\pm$        0.022 &   11.898 $\pm$       0.019 &10.520$\pm$     0.030&   9.722$\pm$     0.027   &8.155$\pm$     0.052&   4.554$\pm$     0.112&WN6\\
WR1667-D00&  19 20 50.49& +13 18 41.1&14.687 $\pm$       0.073&    12.951$\pm$        0.069 &   11.947 $\pm$       0.055&10.569 $\pm$    0.027 & 10.121 $\pm$    0.023  &10.593 $\pm$    0.396 &  7.794 $\pm$    0.323&WN6\\
WR1361-1583& 18 07 05.16& -20 15 16.1& 14.867 $\pm$       0.044  &  12.671 $\pm$       0.043  &  11.162   $\pm$     0.029&10.043  $\pm$   0.037&   9.385$\pm$     0.030&   8.820$\pm$     0.161&   5.351$\pm$     0.108&WN9\\
\hline
\hline
WR1327-14AF\tablenotemark{a}&17 59 02.27  &-24 17 00.1 & 14.401  $\pm$      0.043  &  12.238  $\pm$      0.046  &  10.875    $\pm$     null&10.066  $\pm$     0.046&   9.525 $\pm$      0.035   &9.864  $\pm$      null&   6.536  $\pm$      null&WC8\\
\hline
\hline
WR1627-F10   & 19 10 45.65 & +08 49 27.4 & 15.331    $\pm$    0.048   & 13.660   $\pm$     0.058   & 12.655   $\pm$     0.041 & 11.667   $\pm$  0.031 & 11.038   $\pm$  0.028  & 9.548  $\pm$   0.102 &  6.835 $\pm$    0.101   & Emmision \\
WR1373-1FBA &  18 10 27.36 & -18 56 18.4 & 14.988  $\pm$      0.080  &  12.928   $\pm$     0.056  &  11.401   $\pm$     0.045 & 9.379 $\pm$    0.030  &  8.729 $\pm$    0.026  & 8.717 $\pm$     -100 &  7.311 $\pm$    0.266  & Giant? \\
\enddata
\tablenotetext{a}{WR1327-14AF did not pass all photometric criterion as it had a null 2MASS K$_{s}$ band detection.  However it fell well within all other parameters and when observed was found to be a WR star. }
\end{deluxetable*}
\clearpage
\end{landscape}

\clearpage
\begin{landscape}
\begin{deluxetable*}{lcccrrrrrrrrrrr}
\label{tab:tab1}
\tabletypesize{\scriptsize}
\tablecaption{ Details of Known WR Spectra Sample\label{SXD-LXD-Targets}}
\tablewidth{0pt}
\tablehead{
\colhead{Name\tablenotemark{a}} &
\colhead{RA} &
\colhead{DEC} &
\colhead{SpT} &
\colhead{$J$} &
\colhead{$H$}&
\colhead{$K_{s}$}&
\colhead{$W1$}&
\colhead{$W2$}&
\colhead{$W3$}&
\colhead{$W4$}&
\colhead{Distance}\\
 &
 &
 &
 &
 \colhead{2MASS}&
 \colhead{2MASS}&
 \colhead{2MASS}&
 \colhead{WISE}&
 \colhead{WISE}&
 \colhead{WISE}&
 \colhead{WISE}&
\colhead{kpc}\\
\colhead{(1)} &
\colhead{(2)} &
\colhead{(3)} &
\colhead{(4)} &
\colhead{(5)} &
\colhead{(6)} &
\colhead{(7)} &
\colhead{(8)} &
\colhead{(9)} &
\colhead{(10)} &
\colhead{(11)} &
\colhead{(12)}\\ 
\\
}
\startdata
WR78	&16 52 19.25	&-41 51 16.2		&WN7h            		&5.441$\pm$	0.015&	5.269$\pm$	0.051&	4.978$\pm$	0.013&	4.654$\pm$	0.078&	3.873$\pm$	0.049&	3.749$\pm$	0.016&	2.980$\pm$	0.027&0.9	\\
WR91	&17 20 22.00	&-38 56 47.0		&WN7             		&9.532$\pm$	0.022&	8.765$\pm$	0.016&	8.199$\pm$	0.020&	7.474$\pm$	0.026&	7.007$\pm$	0.021&	5.875$\pm$	0.058&	2.577$\pm$	0.101&3.0	\\
WR98	&17 37 13.75	&-33 27 55.9		&WN8/WC7         	&8.091$\pm$	0.021&	7.548$\pm$	0.031&	7.045$\pm$	0.015&	6.520$\pm$	0.047&	6.193$\pm$	0.026&	5.464$\pm$	0.016&	4.499$\pm$	0.038&2.9	\\
WR79a	&16 54 58.51	&-41 09 03.1	 	&WN9ha           		&5.148$\pm$	0.029&	5.090$\pm$	0.033&	4.904$\pm$	0.011&	4.612$\pm$	0.090&	4.022$\pm$	0.050&	3.955$\pm$	0.015&	3.196$\pm$	0.020&1.4	\\
WR111	&18 08 28.47	&-21 15 11.2		&WC5             		&7.277$\pm$	0.009&	7.143$\pm$	0.045&	6.510$\pm$	0.009&	6.273$\pm$	0.037&	5.935$\pm$	0.022&	5.259$\pm$	0.021&	4.388$\pm$	0.054&1.6\\
WR90	&17 19 29.90	&-45 38 23.8		&WC7             		&6.251$\pm$	0.017&	6.090$\pm$	0.047&	5.520$\pm$	0.015&	5.328$\pm$	0.060&	4.561$\pm$	0.034&	4.238$\pm$	0.015&	3.497$\pm$	0.023&1.0\\
WR79	&16 54 19.70	&-41 49 11.5		&WC7+O5-8     		&5.963$\pm$	0.009&	5.808$\pm$	0.041&	5.389$\pm$	0.019&	5.287$\pm$	0.057&	4.520$\pm$	0.042&	4.169$\pm$	0.016&	3.412$\pm$	0.029&1.2	\\
WR86	&17 18 23.06	&-34 24 30.6		&WC7 (+B0III-I)		&7.436$\pm$	0.017&	7.136$\pm$	0.023&	6.666$\pm$	0.013&	6.369$\pm$	0.033&	5.917$\pm$	0.021&	5.390$\pm$	0.016&	4.723$\pm$	0.032&1.8	\\
WR92 	&17 25 23.15	&-43 29 31.9		&WC9 			&9.503$\pm$	0.021&	9.222$\pm$	0.024&	8.822$\pm$	0.021&	8.296$\pm$	0.023&	7.883$\pm$	0.019&	7.115$\pm$	0.018&	6.450$\pm$	0.071&3.8\\
WR121	&18 44 13.15	&-03 47 57.8		&WC9dusty			&8.295$\pm$	0.021&	7.056$\pm$	0.041&	5.773$\pm$	0.017&	4.852$\pm$	0.091&	3.692$\pm$	0.063&	3.435$\pm$	0.025&	2.974$\pm$	0.028&1.8\\
\hline
\hline
\enddata
\tablenotetext{a}{Sources were obtained from the \citet{van-der-Hucht06} compilation of WR stars}
\end{deluxetable*}
\clearpage
\end{landscape}

\begin{deluxetable*}{ccccccccccccccr}
\label{tab:tab1}
\tabletypesize{\scriptsize}
\tablecaption{Mean NIR and Mid-IR Colors of WN Stars\label{meancolors}}
\tablewidth{0pt}
\tablehead{
\colhead{SpT} &
\colhead{N$_{W1-W2}$\tablenotemark{a}} &
\colhead{N$_{(J-K_{s})}$\tablenotemark{a}} &
\colhead{N$_{(H-K_{s})}$\tablenotemark{a}} &
\colhead{(J-K$_{s}$)$_{avg}$} &
\colhead{$\sigma$(J-K$_{s}$)} &
\colhead{(W1-W2)$_{avg}$} &
\colhead{$\sigma$(W1-W2)} &
\colhead{(H-K$_{s}$)$_{avg}$} &
\colhead{$\sigma$(H-K$_{s}$)} \\
\colhead{(1)}  &
\colhead{(2)}  &
\colhead{(3)}  &
\colhead{(4)}  &
\colhead{(5)}  &
\colhead{(6)}  &
\colhead{(7)}  &
\colhead{(8)}  &
\colhead{(9)}  &
\colhead{(10)}  \\
}
\startdata
  WN3    &    4    &    4       &    4     &      0.36    &      0.10    &      0.28    &      0.03    &      0.22    &      0.07   \\
  WN4    &   23    &   23    &   23    &      1.33    &      1.00    &      0.41    &      0.13    &      0.58    &      0.35   \\
  WN5    &   39    &   35    &   37    &      1.45    &      1.07    &      0.39    &      0.19    &      0.65    &      0.51   \\
  WN6    &   43    &   41    &   41    &      1.49    &      0.92    &      0.41    &      0.16    &      0.60    &      0.32   \\
  WN7    &   51    &   44    &   47    &      1.96    &      1.21    &      0.50    &      0.36    &      0.81    &      0.47   \\
  WN8    &   32    &   27    &   31    &      2.06    &      1.62    &      0.61    &      0.49    &      0.96    &      0.67   \\
  WN9    &   17    &   14    &   17    &      3.12    &      2.34    &      0.55    &      0.34    &      1.21    &      0.95   \\
\hline
\enddata
\tablenotetext{a}{Dusty WC sources were excluded.}
\end{deluxetable*}

\begin{deluxetable*}{ccccccccccccccr}
\label{tab:tab1}
\tabletypesize{\scriptsize}
\tablecaption{Mean NIR and Mid-IR Colors of WC Stars\label{meancolors2}}
\tablewidth{0pt}
\tablehead{
\colhead{SpT} &
\colhead{N$_{W1-W2}$\tablenotemark{a}} &
\colhead{N$_{(J-K_{s}})$\tablenotemark{a}} &
\colhead{N$_{(H-K_{s}})$\tablenotemark{a}} &
\colhead{N$_{Dusty}$} &
\colhead{(J-K$_{s}$)$_{avg}$} &
\colhead{$\sigma$(J-K$_{s}$)} &
\colhead{(W1-W2)$_{avg}$} &
\colhead{$\sigma$(W1-W2)} &
\colhead{(H-K$_{s}$)$_{avg}$} &
\colhead{$\sigma$(H-K$_{s}$)} \\
\colhead{(1)}  &
\colhead{(2)}  &
\colhead{(3)}  &
\colhead{(4)}  &
\colhead{(5)}  &
\colhead{(6)}  &
\colhead{(7)}  &
\colhead{(8)}  &
\colhead{(9)}  &
\colhead{(10)}  &
\colhead{(11)}  \\
}
\startdata
  WC4    &    3    &    3    &    3    &    1    &      1.22    &      0.43    &      0.40    &      0.05    &      0.73    &      0.14   \\
  WC5    &   16    &   16    &   16    &    1    &      1.48    &      0.74    &      0.38    &      0.09    &      0.78    &      0.26   \\
  WC6    &   15    &   14    &   14    &    1    &      1.34    &      0.68    &      0.42    &      0.12    &      0.75    &      0.26   \\
  WC7    &   24    &   20    &   23    &    3    &      1.80    &      1.11    &      0.52    &      0.19    &      0.96    &      0.47   \\
  WC8    &   43    &   35    &   40    &    7    &      3.02    &      1.41    &      0.60    &      0.15    &      1.29    &      0.48   \\
  WC9    &   18    &   18    &   19    &   22    &      3.16    &      2.65    &      0.63    &      0.34    &      1.28    &      0.86   \\
\hline
\enddata
\tablenotetext{a}{Dusty WC sources were excluded.}
\end{deluxetable*}

\clearpage
\end{document}